\documentclass[twoside,11pt]{article}

\usepackage{microtype}
\usepackage{graphicx}
\usepackage{subfigure}
\usepackage{booktabs} 

\usepackage{amsmath}
\usepackage{amsfonts}
\usepackage{mathtools}
\usepackage{bbm}
\usepackage{multirow}
\usepackage{enumitem}
\usepackage{multicol}
\usepackage{xspace}
\RequirePackage{algorithm}
\RequirePackage{algorithmic}

\newtheorem{exmp}{Example}[section]

\newcommand{\ybold}{\boldsymbol{y}}
\newcommand{\xbold}{\boldsymbol{x}}

\newcommand{\betabold}{\boldsymbol{\beta}}
\newcommand{\epsilonbold}{\boldsymbol{\epsilon}}
\newcommand{\zbold}{\boldsymbol{z}}
\newcommand{\bbold}{\boldsymbol{b}}

\newcommand{\E}{\mathbb{E}}

\newcommand{\Var}{\mathrm{Var}}
\newcommand{\Cov}{\mathrm{Cov}}
\newcommand{\tr}{\mathrm{tr}}
\newcommand{\argmin}{\mathrm{argmin}}

\newcommand{\alname}{RETCO\xspace}

\usepackage{jmlr2e}

\ShortHeadings{Trees-Based Models for Correlated Data}{Rabinowicz and Rosset}
\firstpageno{1}

\begin{document}

\title{Trees-Based Models for Correlated Data}

\author{\name Assaf Rabinowicz \email assafrab@gmail.com \\
       \addr Department of Statistics and Operations Research\\
       Tel Aviv University\\
       Tel Aviv, Israel
       \AND
       \name Saharon Rosset \email saharon@tauex.tau.ac.il \\
       \addr Department of Statistics and Operations Research\\
       Tel Aviv University\\
       Tel Aviv, Israel}


\maketitle

\begin{abstract}
This paper presents a new approach for regression trees-based models, such as simple regression tree, random forest and gradient boosting, in settings involving correlated data. We show the problems that arise when implementing standard regression trees-based models, which ignore the correlation structure. Our new approach explicitly takes the correlation structure into account in the splitting criterion, stopping rules and fitted values in the leaves, which induces some major modifications of standard methodology. The superiority of our new approach over trees-based models that do not account for the correlation is supported by simulation experiments and real data analyses.
\end{abstract}


\section{Introduction}
Trees-based models are widely used for tabular data due to their high prediction accuracy and their inherent model selection functionality \citep{Hastie2009elements}.
Commonly, trees-based models are fitted without assuming any distributional setting on the dependent variable. While the distribution of the dependent variable is mostly unknown and therefore it is tempting to avoid distributional assumptions, the correlation structure, which relates to the sampling mechanism (e.g., clustered data, time-series data, longitudinal data, spatial data), is frequently known and therefore it is not reasonable to ignore it. Unlike in trees-based models, the correlation structure is an essential component in many machine learning models, for example kernel covariance functions are used in Gaussian processes regression \citep{rasmussen2003gaussian}, which is frequently implemented for modeling data sets with spatial correlation structure, such as  neuroscience data sets \citep{caywood2017gaussian} and climatography data sets \citep{goovaerts1999geostatistics}. Another example is linear mixed model, which is used for data involving longitudinal correlation structure, as is common in health \citep{coull2001respiratory} and trading \citep{westveld2011mixed} applications.

In this paper we develop a method which combines the concepts of \textit{random effects} and \textit{random fields}, which are convenient platforms for analyzing correlated data, and trees-based models such as: regression tree, random forest and gradient boosting. The desired result is that the trees-based part results a high prediction accuracy and model selection capabilities and the random effects part enables to boost the model performance by using the correlation structure and even to exract statistical inference. The idea of integrating between random effects/random field and trees-based methods has previously been explored (see \citealt{sela2012re,stephan2015random}, discussed in more detail in Section \ref{Section: Comparison to Other Methods}). 
However, we propose a novel approach which takes advantage of recent developments in model evaluation and selection methodologies for correlated settings, and yields improved results as demonstrated below. 

Section \ref{Section: Theoretical Background} gives relevant background for the proposed method. The background contains a brief description of trees-based methods, linear mixed model (which is based on random effects) and prediction error estimation for correlated data, which has a key role in our approach. Our new algorithm, REgression Tree for COorrelated data (\alname), is presented and discussed in Section \ref{Section: Trees-Based Linear Mixed Models}. Section \ref{Section: Comparison to Other Methods} compares \alname with other algorithms that were introduced in recent years. Section \ref{Section: Numerical Results} presents simulation and real data analyses that support our proposed algorithm.
\section{Theoretical Background}\label{Section: Theoretical Background}
This section presents briefly regression trees-based models, linear mixed models and prediction error estimation for correlated data. Additional information can be found in Appendix \ref{App: Theoretical Background} and in resources which are cited below.

\subsection{Trees-Based Models}\label{Section: Trees-Based Models}
Given a vector of covariates $\xbold^*\in\mathbb{R}^{p}$ a regression tree estimates the corresponding response, $y^*\in \mathbb{R},$ as follows: \[f(\xbold^*)=\sum_{s=1}^{S} I_{(\xbold^*\in g_{s})}\mu_{s},\]
where
$g_{s}\subseteq\mathbb{R}^p$ and $\mu_{s}\in\mathbb{R},$ for all $s\in\{1,...,S\}.$ $\{g_s\}_{s=1}^S$ define a partition of the covariate space, that is $g_s\cap g_t = \emptyset$ for $s\neq t$ and $\cup_{s=1}^S g_s$ is the entire covariates space. $\mu_{s}$ is the predictor for covariate vectors that are in $g_s.$ The nodes, $\mathcal{S}=\{\mu_{s},g_{s}\}_{s=1}^{S}$ are selected using a recursive optimization process that can be diagrammed as a tree, where $\mathcal{S}$ are the terminal nodes (the leaves). The recursive optimization for selecting 
$\mathcal{S}$ is based on the training set, $\{y_i,\xbold_i\}_{i=1}^n=\{\ybold,X\},$ where $\ybold\in\mathbb{R}^n$ and $X\in\mathbb{R}^{n\times p}$ is the design matrix, and it is implicitly assumed that the prediction points, $y^*$ and $\xbold^*,$ are drawn from the same distribution as $y_i$ and $\xbold_i.$ Each step in the tree's recursive optimization process is a model selection of linear models, where a threshold of one of the available covariates is selected in order to minimize a loss function, $Loss(\cdot,\cdot):\mathbb{R}^2\to\mathbb{R}.$ Also, stopping rules that follow predefined hyper-parameters, e.g., maximal depth of the tree and minimal number of training set observations in a node, are enforced on the recursive optimization and limit the tree's depth and affect other properties of the tree's structure in order to avoid overfitting. 

Random forest (RF) and gradient boosting (GB) predictor are based on averaging an ensemble of trees. More information about regression tree, RF and GB, as well a formalized regression tree algorithm, is available in Appendix \ref{App: Theoretical Background} and can also be found in \citet{freund1999short,friedman2001greedy,breiman2001random,Hastie2009elements}.

\subsection{Linear Mixed Model}\label{Section: Linear Mixed Models}
In linear mixed models (LMM) there are two covariate vectors: fixed effects covariates, $\xbold^*\in \mathbb{R}^{p},$ and random effects covariates, $\zbold^*\in \mathbb{R}^{q}.$ Commonly, $y^*$ is assumed to be normally distributed and decomposed as follows:
\begin{align*}
y^*=\betabold^{t}\xbold^{*}+\bbold^{t}\zbold^{*}+\epsilon^*,
\end{align*}
where $\betabold$ is the fixed effects vector of coefficients, $\bbold\sim N_q(0,G)$ is the random effects vector and $\epsilon^*\sim N(0,\sigma^2)$ is the residual. $\betabold^{t}\xbold^*$ is the \textit{marginal mean} of $y^*,$ $\E(y^*|\xbold^*),$ and $\betabold^{t}\xbold^*+\bbold^{t}\zbold^*$ is the \textit{conditional mean} of $y^*$ given $\bbold,$ which is commonly denoted as $\E(y^*|\xbold^*,\zbold^*,\bbold).$

$\betabold$ and $\bbold$ are estimated using the training sample, $\{y_i,\xbold_i,\zbold_i\}_{i=1}^n=\{\ybold,X,Z\},$ which follows the same model: 
\[\ybold=X\betabold+Z\bbold+\epsilonbold,\]
where $\epsilonbold\sim N_n(0,\sigma^2I_n),$
and $X,Z$ are the fixed effects and random effects covariate matrices, respectively.
Since $y^*$ and $\ybold$ share the same random effects, $\bbold,$ they are correlated, and estimating $\bbold$ by the training sample can later be used for improving the prediction accuracy of $y^*.$ Given $G$ and $V\coloneqq\Var(\ybold)=ZGZ^{t}+\sigma^2I_n,$ $\betabold$ and $\bbold$ can be estimated as follows:
\begin{align*}
\widehat{\betabold}&=(X^{t}V^{-1}X)^{-1}X^{t}V^{-1}\ybold\\
\widehat{\bbold}&=GZV^{-1}(\ybold-X\widehat{\betabold}).
\end{align*}
\citet{harville1976extension} showed that given the true covariance matrices, $G$ and $V,$ the estimated \textit{conditional mean}, $\widehat{\E}(y^*|\xbold^*,\zbold^*,\widehat{\bbold})=\widehat{\betabold}^{t}\xbold^*+\widehat{\bbold}^{t}\zbold^*,$ is the best linear unbiased predictor (BLUP) of $y^*.$ In practice, the covariance matrices are mostly unknown and therefore are estimated using maximum likelihood or restricted maximum likelihood \citep{verbeke1997linear}. Note that given the covariance matrices, LMM is linear in $\ybold,$ i.e., the LMM predictor satisfies:
\[\widehat{y}^*=h^*\ybold,\]
where $h^*,$ the hat vector, does not contain the training response vector $\ybold,$ and the element $h^*[i]$ is the \textit{weight} of $y_i$ in predicting $y^*,\forall i\in[1,...,n].$
\subsubsection{\texorpdfstring{$\bbold^*\neq\bbold$}{TEXT}\label{new random effects}
Scenario}
In many cases the random effects of $y^*$ are not the same as the random effects of $\ybold,$
i.e.,
\[
y^*=\betabold^{t}\xbold^*+\bbold^{*{t}}\zbold^*+\epsilon^*,\;\;\bbold^*\neq\bbold.
\]
That means that the correlation between the observations in $\ybold$ is not the same as the correlation between $y^*$ and the observations in $\ybold.$ In case $\bbold^*\perp\bbold,$ which implies $\Cov(y^*,\ybold)=0,$ estimating $\bbold$ does not improve the prediction accuracy of $y^*.$ Therefore, in this case $y^*$ is predicted by the \textit{marginal mean} $,\widehat{\E}(y^*|\xbold^*)=\widehat{\betabold}^{t}\xbold^*.$ This model, which is a special case of LMM, is also called generalized least squares model (GLS). A simple example for this scenario, is when the prediction set contains different clusters than in the training set. Several prediction tasks that follow this scenario setting are analyzed in Section \ref{Real Data Results}. For example, using FIFA data set from \href{https://www.kaggle.com/}{Kaggle website}, a predictive model for footballs players' market-value was trained using GLS, where the prediction goal is to predict the market-value of players that belong to clubs that \textit{do not appear} in the training set. This data set has a clustered correlation structure, where cluster is the players' club, i.e., market-values of players that belong to the same club are correlated, while market-values of players that belong to different clubs are uncorrelated. Given the prediction goal of predicting the market-value of players from new clubs, this setting follows exactly the $\bbold^*\perp\bbold$ setting.

Another scenario is when $\bbold^*\not\perp\bbold$ (although $\bbold^*\neq\bbold$) and therefore $\Cov(y^*,\ybold)\neq0.$ This can happen for example when some of the random effects of $\ybold$ and $y^*$ are the same and some are not. In this case, the elements in $\bbold$ that are in $\bbold^*$ should be estimated and used for predicting $y^*.$ Examples where $\bbold^*\not\perp\bbold$ although $\bbold^*\neq\bbold,$ and more information about this scenario can be found in \cite{rabinowicz2020cross}. This scenario, of $\bbold^*\neq\bbold,$ is common and should be taken into account when developing trees-based methods for correlated data, as will be developed in Section \ref{Section: Trees-Based Linear Mixed Models}.
 
\subsection{Prediction Error Estimation and Model Selection for Correlated Data}\label{Section: Prediction Error Estimation and Model Selection for Correlated Data}
Once a predictive model is fitted, it is often evaluated by its prediction error estimator. Moreover, when there is a set of alternative models (e.g., for LMM: models with different covariates, for trees-based models: models with different hyper-parameters) the 'best' model can be selected based on minimizing the prediction error estimator. It is important to note that common prediction error estimators, e.g.,  AIC \citep{akaike1974new}, Cp \citep{mallows1973some}, and even cross-validation \citep[CV]{stone1974cross} are biased in some settings involving correlated data. Naturally, their corresponding model selection criteria are also biased in those scenarios. This bias was studied in the recent years, mostly for linear models. Here we present Cp, AIC and CV versions for correlated data. For description of the original Cp and AIC versions, that do not address correlation structure, see Appendix \ref{App: Theoretical Background}. 

\subsubsection{Cp}
In Cp, the goal is to estimate the squared prediction error:
\[\small\E_{\ybold,\ybold^*}\frac{1}{n}\|\ybold^*-H\ybold\|^2_2,\]
where $H$ is the hat matrix, and $\ybold^*\in\mathbb{R}^n$ is a vector of new observations measured at the same covariate values as $\ybold,$ $\{X,Z\},$ but with new independent noise and potentially different random effects realizations. This type of prediction error, when both $\ybold$ and $\ybold^*$ relate to the same covariate points, $\{X,Z\},$ is called \textit{in-sample prediction error}. In this setting, it is natural to consider $\{X,Z\}$ as fixed  matrices rather than random variables. \citet{hodges2001counting} extended Cp to LMM
with $\bbold^*=\bbold,$ here we employ a more general formulation which reduces to \citet{hodges2001counting} when $\bbold^*=\bbold$ but also covers the case that they are different:
\begin{align*}\small
Cp=\frac{1}{n}\|\ybold-\widehat{\ybold}\|_2^2+\frac{2}{n}\tr\Big(H\big(\Var(\ybold)-\Cov(\ybold^*,\ybold)\big)\Big).    
\end{align*}
\subsubsection{AIC}
Similarly to Cp, AIC is also an in-sample error, however its loss function is based on likelihood. \citet{vaida2005conditional} presented the conditional AIC (cAIC) and marginal AIC (mAIC) which are suitable for the scenarios where $\bbold^*=\bbold$ and $\bbold^*\perp\bbold,$ respectively. Here we will use the name AIC for our formulation which subsumes cAIC and mAIC, but also covers the $\bbold^*\neq\bbold\cap\bbold^*\not\perp\bbold$ scenario: 
\begin{align*}\small
  AIC=&-\frac{2\ell\big(\ybold;\widehat{\E}(\ybold|X,Z,\widehat{\bbold}),V_c\big)}{n}+\frac{2\tr\Big(H\big(\Var(\ybold)-\Cov(\ybold,\ybold^*)\big)V^{-1}_c\Big)}{n},  
\end{align*}
where $\ell(\ybold;\widehat{\E}(\ybold|X,Z,\widehat{\bbold}),V_c)$ is the conditional likelihood of $\ybold$ given $\widehat{\bbold},$ and $V_{c}=\Var(\ybold^*|\bbold).$
\subsubsection{Cross-Validation (CV)} Unlike Cp and AIC, CV estimates the \textit{generalization error}:
\[
\E_{X,X^*,Z,Z^*}\E_{\ybold^*,\ybold}\sum_{i=1}^{n}\frac{1}{n}Loss\big(y^*_i,\widehat{y}_i^*(\xbold_i^*,\zbold_i^*;\ybold,X,Z)\big),
\]
where $\ybold^*$ is drawn from the same marginal distribution as $\ybold,$ but relates to \textit{new covariate values}, $\{\xbold^*_i,\zbold^*_i\}_{i=1}^n=\{X^*,Z^*\},$ which were drawn from the same distribution as $\{X,Z\}.$ For simplicity, a special case of CV algorithm, leave-one-out (LOO), is presented:
    \begin{enumerate}
        \item $\forall i\in [1,...,n],$ fit a model using the whole sample besides the $i^{th}$ observation. Denote the sample without the $i^{th}$ observation as $\{\ybold_{-i},X_{-i},Z_{-i}\}.$ 
        \item Predict $y_i$ by the fitted model and denote the predictor as $\widehat{y}_i^{-i}=\widehat{y}_i(\xbold_i,\zbold_i;\ybold_{-i},X_{-i},Z_{-i}).$ \end{enumerate}
For a squared errors loss function and linear predictor of $\ybold,$  the CV error is
\[CV=\frac{1}{n}\|\ybold-H_{cv}\ybold\|_2^2,\]
where $H_{cv},$ the CV hat matrix, is
\[
H_{cv}=\left[\begin{array}{cccc}
0 & h_{1,2} & ... & h_{1,n}\\
h_{2,1} & 0 &  & h_{2,n}\\
...\\
h_{n,1} & h_{n,2} & ... & 0
\end{array}\right],
\]
$h_{k,k'}\in\mathbb{R}\;\forall k,k'\in \{1,...,n\}.$ In this presentation, the vector $[h_{1,2},...,h_{1,n}]$ is the hat matrix of $\widehat{y}_1^{-1}.$ K-fold CV generalizes LOO, by partitioning $\{\ybold,X,Z\}$ into K equal size subsets, $\{\ybold_k,X_k,Z_k\}_{k=1}^K,$ where $K\leq n$ (rather than $K=n$ as in LOO)

\citet{rabinowicz2020cross} presented a generalization of CV, $CV_c,$ which is suitable for scenarios involving correlated data:
\begin{align*}\small
CV_c=CV+\frac{2}{n}\tr\Big(H_{cv}\big(\Var(\ybold)-\Cov(\ybold^*,\ybold)\big)\Big).
\end{align*}
In the LMM settings, when $\bbold^*=\bbold,$ $CV_c=CV.$ For more information see \citet{rabinowicz2020cross}.

Cp and $CV_c$ do not assume a specific distributional setting and can be applied for any linear model, while AIC is suitable for LMM, however it can also be adjusted for other linear models that assume normality, such as Gaussian process regression (GPR). Table \ref{Table: Prediction Error Summary} summarizes the prediction error estimators that are described in this section.
\begin{table}
\begin{center}
\begin{small}
\begin{tabular}{|lcc|}
\toprule
Method& Prediction Error Type&Distributional Assumptions \\
\midrule
Cp   & in-sample error & -- \\ \hline
AIC &  in-sample error& normal likelihood \\\hline
$CV_c$ & generalization error &--\\ 
\bottomrule
\end{tabular}
\end{small}
\end{center}
\caption{Summary of prediction errors for correlated data}\label{Table: Prediction Error Summary}
\end{table}
\section{Trees-Based Models for Correlated Data}\label{Section: Trees-Based Linear Mixed Models}
This section presents the main algorithm of this paper, REgression Tree for COrrelated data (\alname), and discusses the main differences between \alname and the standard regression tree algorithm.
\subsection{\alname Algorithm}\label{section: RETCO algorithm}
A simple approach for integrating between trees-based methods and random effects is replacing the marginal mean in LMM, $\betabold^{t}\xbold^*,$ by a trees-based model, $f(\xbold^*):$
\begin{align}
f(\xbold^*)+\bbold^{t}\zbold^*\label{BLUP with tree},
\end{align}
where $f(\xbold^*)$ is created in a way that does account for the correlation structure (unlike in the standard regression tree algorithm). The power of the model in expression (\ref{BLUP with tree}) can be perceived from different points of views. From the LMM point of view, the additive representation of marginal and conditional means is preserved, however the marginal mean is non-linear and therefore more expressive than in standard LMM. From the regression tree point of view, this approach differentiates between the two types of covariates---fixed effects, which are used for splitting the tree's nodes, and random effects that are added linearly to the fitted tree---enabling expressing and using the correlation structure. This also enables using inference tools that do not exist for standard regression trees but exists in LMM, for example, comparing between the variance components, $\sigma^2$ and $G.$ Note that expression (\ref{BLUP with tree}) assumes that the effect of the random effects on $\ybold$ is linear, however it can be generalized.

The template in expression (\ref{BLUP with tree}) was already suggested (for literature review see Section \ref{Section: Comparison to Other Methods}), however here we present a new algorithm, \alname, for fitting $f(\xbold^*)$ that follows the theoretical aspects that were presented in the previous section. The algorithm formulation is general, and is not based on a specific prediction error type or distributional setting. Also, the algorithm refers to a case when $f(\cdot)$ is a single regression tree, the extension to RF and GB will be discussed in Section \ref{Section: Extension for RF and GB}. 

Algorithm \ref{algorithm}, which presents \alname use the following notations:
\begin{itemize}
    \item $f(\xbold_i|\mathcal{S}),$ the intermediate predictor of $y_i$ during the tree fitting. 
    \[f(\xbold_i|\mathcal{S})=\sum_{s\in\mathcal{S}}I_{(\xbold_i\in g_l)}\eta(\mu_s),\]
    where 
    \begin{itemize}
        \item $\mu_s$ is the  GLS predictor of $\{y_i|\xbold_i\in g_s\},\;\forall s\in\mathcal{S}$ \item   $\eta(\cdot)$ is the identify function for the setting of  $\bbold^*\perp\bbold,$ and the BLUP for  $\bbold^*\not\perp\bbold.$ 
    \end{itemize}
    Of course, after fitting the tree $f(\xbold_i)=f(\xbold_i|\mathcal{S}).$
\item 
$f\left(\xbold_i|j,c,\mathcal{S}/s\right),$ the predictor of $y_i$ when splitting node $s$ using covariate $j$ at the threshold $c:$
 \[f\left(\xbold_i|j,c,\mathcal{S}/s\right)= I_{(\xbold_i\in g_s\;\cap \;x_{i,j}\leq c)}\eta\big(\mu_s^l(c)\big)+I_{(\xbold_i\in g_s\;\cap \;x_{i,j}>c)}\eta\big(\mu_s^r(c)\big)+\sum_{l\in\mathcal{S}/s}I_{(\xbold_i\in g_l)}\eta(\mu_l),
     \]
     where $\mu_s^l(c)$ and $\mu_s^r(c)$ are the GLS predictors of 
 $\{y_i|\xbold_i\in g_s\;\cap \;x_{i,j}\leq c\}$ and $\{y_i|\xbold_i\in g_s\;\cap \;x_{i,j}> c\},$ respectively. 
\end{itemize}
\begin{algorithm}
\caption{REgression Tree for COrrelated Data (\alname)}
\label{algorithm}
\begin{algorithmic}
\STATE {\bfseries Input:} $\ybold,\;X,\;Z.$ 
\STATE {\bfseries Output:} $f(\cdot).$ 
 \STATE {\bfseries High level setting:}  \label{Algo: error selection}
 \begin{itemize}
     \item select a prediction error estimator loss function---Cp, AIC or $CV_c$---from Table \ref{Table: Prediction Error Summary}
     \item define the relation between $\bbold^*$ and $\bbold$ ($\bbold^*=\bbold$ versus $\bbold^*\neq\bbold$)
     \item define the stopping rules
 \end{itemize}
 \STATE  {\bfseries Initialization:} \label{Algo: initialize}
 $\mathcal{S}=\{g_1,\mu_1\},$
where $g_1=\mathbb{R}^p$ and
$\mu_1$ is the GLS estimator.
 \REPEAT
 \STATE   Given the predefined stopping rules, find the best node for splitting ($\tilde{s}$), the best covariate ($j_{\tilde{s}}$) and the best threshold ($c_{{\tilde{s}}}$) as follows:
 \begin{align}\nonumber
 \tilde{s},j_{\tilde{s}},c_{{\tilde{s}}}&=\underset{s\in\mathcal{S},j\in J_s,c\in\mathbb{R}}{\argmin}\frac{1}{n}\sum_{i=1}^{n}Loss\big(y_i,f\left(\xbold_i|j,c,\mathcal{S}/s\right)\big)\\\label{algo: stopping rule}
 s.t:&\;\;\frac{1}{n}\sum_{i=1}^{n}Loss\big(y_i,f\left(\xbold_i|j,c,\mathcal{S}/s\right)\big)<\frac{1}{n}\sum_{i=1}^{n}Loss\big(y_i,f\left(\xbold_i|\mathcal{S}\right)\big),\\\nonumber \end{align}
 where $J_s$ is the set of available covariates for splitting node $s.$
 \IF {$\{\tilde{s},j_{\tilde{s}},c_{\tilde{s}}\}$ exist}
 \STATE
     Update $\mathcal{S}:$ replace $(g_{\tilde{s}},\mu_{\tilde{s}})$ by the new two nodes, $\big(g_{\tilde{s}}\cap x_{j_{\tilde{s}}}\leq c_{\tilde{s}},\mu_{\tilde{s}}^r(c_{\tilde{s}})\big),$ $\big(g_{\tilde{s}}\cap x_{j_{\tilde{s}}}> c_{\tilde{s}},\mu_{\tilde{s}}^l(c_{\tilde{s}})\big),$ where $x_{j_{\tilde{s}}}$ is the covariate $j_{\tilde{s}}.$\label{Algo: update the nodes}
     \ENDIF
 \UNTIL {$\{\tilde{s},j_{\tilde{s}},c_{\tilde{s}}\}$ do not exist or stopping rules are satisfied $\forall s\in\mathcal{S}.$}
\end{algorithmic}
\end{algorithm}

As can be seen in Algorithm \ref{algorithm}, \alname covers various settings including different prediction error measures and correlation structures. Before analyzing its properties, here are some technical details for \alname:\\
\begin{itemize}
\item Predefined stopping rules: stopping rules, such as maximal tree's depth and minimal number of observations in a node, are commonly applied when fitting regression trees (for more details, see Section \ref{Section: Trees-Based Models}).
    \item Variance estimation: the variance components, needed for calculating $f\left(\xbold_i|j,c,\mathcal{S}/s\right),$ $f\left(\xbold_i|\mathcal{S}\right)$ and the loss function, can be estimated in different ways, for example using maximum likelihood or restricted maximum likelihood. For clustered data, simple closed-form equations that estimate the variance components are available and presented in Appendix \ref{App: Estimating variance components for the random intercept model}.
\item Main optimization part: 
    \begin{itemize}
        \item  the GLS predictors, $\mu_s^l(c),\;\mu_s^r(c)$ and $\{\mu_l\}_{l\in\mathcal{S}/s}$ are estimated using dummy variables for the current leaves: $\{i|\xbold_i\in g_s\;\cap \;x_{i,j}\leq c\},$ $\{i|\xbold_i\in g_s\;\cap \;x_{i,j}> c\}$ and  $\{i|\xbold_i\in g_l\}_{l\in\mathcal{S}/s}$
        \item for every potential split, the variance components are estimated using the whole sample (rather than only $\{y_i|\xbold_i\in g_s\}$)
        \item the condition of eq. (\ref{algo: stopping rule}) is required since the loss function is not a training error (as in the standard regression tree algorithm) and therefore splitting a node may increase the loss
    \end{itemize}
\item Algorithm's Output: since the tree estimates the marginal mean, its predictors are $\{\mu_l\}_{l\in\mathcal{S}}$ rather than $\{\eta(\mu_l)\}_{l\in\mathcal{S}}.$
\end{itemize}



There are two main conceptual differences between \alname and the standard regression tree algorithm. The first is the use of a prediction error estimator: Cp, AIC or $CV_c$ as loss functions. Unlike in the standard regression tree, where the loss function is the training error, here the loss function is prediction error estimator for correlated data. As was mentioned in Section \ref{Section: Trees-Based Models}, although the regression tree model is a non-linear function of $\ybold,$ each split is a model selection problem of linear models. Due to the linearity, Cp, AIC and $CV_c$ can be implement. More details about the effect of using these prediction error estimators on the selected thresholds and variables are given in Section \ref{Section: The Correlation Penalty Effect}. The second conceptual difference is the \textit{iterative} approach that is used here instead of the recursive approach that is used in the standard regression tree model. The iterative approach is expressed by selecting the optimal node, $\tilde{s},$ for splitting, rather than splitting each node independently. The reason for using an iterative approach is that observations in different paths are dependent. The dependency is accounted in the GLS predictors as well as in other expressions in the loss function (e.g., bias correction and likelihood, depending on the correlation setting and the selected loss function), therefore splitting one node may affect the splitting of the other. This is in contrast to i.i.d setting with training error loss function, where a recursive approach can be used since observations in different paths are independent and therefore splitting one node does not affect the splitting of the other. 

For LMM-like setting, when $\bbold^*=\bbold,$ once $f(\cdot)$ was fitted, the random effects can be estimated using the BLUP formula:
\[\widehat{\bbold}=\widehat{\Cov}(\ybold^*,\ybold)\widehat{V}^{-1}\big(\ybold-f(X)\big),\]
where $f(X)=[f(\xbold_1),...,f(\xbold_n)].$  

\alname is formalized in context of LMM: $\{\mu_s\}_{s=1}^S,$ are GLS predictors and the variance is decomposed as in LMM, $V=ZGZ^{t}+\sigma^2\times I_n$). However these properties are not fundamental in the algorithm and can easily be generalized. For example, the covariance matrices, $\Cov(\ybold^*,\ybold)$ and $\Var(\ybold),$ can be expressed using a kernel covariance function as is common in Gaussian process regression (and will be analyzed numerically in Section \ref{Section: Numerical Results}). Also, other linear models instead of GLS can be used for estimating the marginal means, $\{\mu_s\}_{s=1}^S.$ Moreover, the random effects, $\bbold,$ can be estimated in different ways than using the BLUP formula. 

Due to the iterative approach the complexity is higher than the complexity of a standard regression tree. Assuming $x_{i,j}\neq x_{i',j},\forall i\neq i'\in[1,...,n],\;j\in[1,...,p],$ the loss function is evaluated less than $n\times p\times 2^d$ times, where $d$ is the tree's depth. Practically, due to hyper-parameters that restrict the number of potential splitting values (e.g., such as minimum observations at each node) and due to the constraint in inequality (\ref{algo: stopping rule}), which frequently shortens the depth of paths, the number of evaluations is smaller. The complexity of each loss function evaluation depends on the loss function, the predictor type and the correlation structure. For example, in Cp loss function, GLS predictor and clustered data, each evaluation complexity is $O(n_c^3),$ where $n_c$ is the cluster size, and therefore the overall computational complexity is $O\big(n\times p\times 2^d\times n_c^3\big).$ 
The amount of memory required is quadratic in $n,$ due to storing of $\Var(\ybold).$ 
\subsubsection{\alname for RF and GB}
\label{Section: Extension for RF and GB}
Extending \alname to RF and GB is done by averaging ensemble of implementations of \alname, while taking into account the special adjustments that RF and GB require. The random effects are estimated in the same way. Several aspects should be noted when implementing RF and GB:
\begin{itemize}
    \item \textit{Training set sampling}: sampling with replacement cannot be implemented naively when the loss function involves calculation of $\Var(\ybold)^{-1}$ (as when GLS predictor or marginal likelihood loss are used)
    since in that case $\Var(\ybold)$ might be a singular matrix due to duplication. Therefore, a half-sample method should be used.
    \item 
    \textit{Number of trees}: since the trees in RF and GB are correlated, then in order to reduce the variance the number of trees should be large, especially when the trees are deep. When correlated data are involved, the trees are even more correlated due to the correlation between the observations. Therefore, the number of trees should be even larger than in the i.i.d sample setting. RF based on \alname is demonstrated in Section \ref{Section: Numerical Results}.
    \item \textit{Response in GB}: in common regression GB the trees are fitted consecutively to the residual of the previous tree. Therefore, the input of the algorithm in all the trees except the first one is not $\ybold$. Correspondingly, the estimated variance matrices relate to the residual of the previous tree, rather than to $\ybold.$ 
\end{itemize}
\subsubsection{Using CV Loss in Regression Tree}
In typical predictive modeling settings, generalization error is the primary objective of learning, hence CV loss is the natural choice. Surprisingly, there is not much previous work on using CV loss in trees, even without correlation. Notable exceptions are the ALOOF algorithm \citep{painsky2016cross} and approximations used in CatBoost \citep{prokhorenkova2017catboost}. The main drawback in using CV-based loss function is increasing the computational cost compares to Cp loss function. Moreover, since \alname is iterative rather than recursive, the number of evaluations of the loss function can remain large for all the splits along the tree.
\subsection{The Bias Correction Effect}\label{Section: The Correlation Penalty Effect}
As explained in the previous sections, we suggest to add a bias correction term to the training error such that the loss function estimates the prediction error. This section illustrates the effect of the bias correction on split selection. Extensive numerical analysis is presented in Section \ref{Section: Numerical Results}.

\subsubsection{\texorpdfstring{$\bbold^*\perp\bbold$}{TEXT} Scenario}\label{Section: The penalty effect, new b}
Observations with positive correlation are similar in higher probability than uncorrelated observations. Therefore, loss functions that do not take into account the correlation, tend to split a node based on the correlation structure of the training set observations rather than their mean. Splitting based on the correlation of the training data is not useful for predicting uncorrelated observations. Therefore, when $\bbold^*\perp\bbold$ it is important to fit the regression tree based on the marginal mean only. The corrections in Cp and $CV_c,$ which take into account the correlation structure, balance this tendency.
Examples \ref{Ex. 1} and \ref{Ex. 2} demonstrate this mechanism. The code for the examples, as well as for the numerical part in Section \ref{Section: Numerical Results}, is written in Python and is available in \url{https://github.com/AssafRab/RETCO}.
\begin{exmp}\label{Ex. 1}
Consider the setting of $\bbold^*\perp\bbold$ and a training data containing four observations from two clusters with the covariance matrix $\small\Var(\ybold)=\scriptsize \begin{pmatrix}2 & 1 & 0 & 0\\1 & 2 & 0 & 0\\0 & 0 & 2 & 1\\0 & 0 & 1 & 2\\\end{pmatrix},$ i.e., observations 1 and 2 belong to the first cluster and observations 3 and 4 belong to the second cluster. The $CV_c$ correction is reduced in this setting to $2\tr\big(H_{cv}\Var(\ybold)\big)/n.$ Two models with the GLS predictor are tested:
\begin{itemize}
    \item Model A, which splits the training set into the two clusters. Given the covariance matrix:
    \[\small H_{cv}= \begin{pmatrix}0 & 1 & 0 & 0\\1 & 0 & 0 & 0\\0 & 0 & 0 & 1\\0 & 0 & 1 & 0\end{pmatrix}\to 
\frac{2}{n}\tr\big(H_{cv}\Var(\ybold)\big)=2.\]
\item Model B, that mixes between the clusters and selects observations 1 and 3 for one subset and 2 and 4 for the other subset. Given the covariance matrix:
\[\small H_{cv}=\begin{pmatrix}0 & .25 & 1 & -.25\\.25 & 0 & -.25 & 1\\1 & -.25 & 0 & .25\\-.25 & 1 & 0.25 & 0\end{pmatrix}\to  \frac{2}{n}\tr\big(H_{cv}\Var(\ybold)\big)=0.5.\]
\end{itemize}

\end{exmp}
As we can see in Example \ref{Ex. 1}, decomposing the penalty, $2\tr\big(H_{cv}\Var(\ybold)\big)/n,$ shows that the weights that relate to observations from the same cluster are multiplied by their positive covariance values and therefore contribute to the penalty, while weights that relate to observations from different clusters are multiplied by zero and therefore do not contribute to the penalty. As a result the penalty of model A, which gives the whole weight for observations from the same cluster, is larger than for model B. Therefore, while $CV$ selects model A when $CV(A)<CV(B),$ $CV_c$ selects model A when  $CV(A)<(CV(B)-1.5).$ In that way $CV_c,$ as well as $Cp,$ balance the tendency to split based on based on the correlation structure of the training set. Obviously, as much much as the observations in the training set are more correlated, the penalty effect is stronger, and the superiority of \alname over the standard algorithm is more prominent (see also Section \ref{Section: Numerical Results}).
\begin{exmp}\label{Ex. 2}
Consider the setting of $\bbold^*\perp\bbold$ and $\ybold\sim N_{100}(0.1\times\xbold_1,V),$
where
\[\small
\Var[i,j]=\begin{cases}
    2, & \text{when } i=j\\
    1, & \text{when } i\neq j \text{ and } (i,j\leq 50\; \text{ or } i,j>50)\\
    0,& \text{ o.w }
    \end{cases},
\]
and
\begin{align*}\small
    \xbold_1[i]&=
    \begin{cases}
    0.5+\epsilon_x, & \text{when i is odd},\quad 
    \epsilon_x\sim N(0,0.1)\\
    -0.5+\epsilon_x, & \text{when i is even},  \end{cases}
\end{align*}
i.e., $\ybold$ contains two clusters of $50$ observations each, and its mean is not correlated with the clusters. 

Two models are tested, model A which uses the threshold  $\xbold_1=0$ and model B which uses $\xbold_2=0,$ where $\xbold_2$ is highly correlated with the clusters:
\begin{align*}
\xbold_2[i]&=\begin{cases}
    0.5+\epsilon_x, & \text{when } i\leq50,\quad \epsilon_x\sim N(0,0.1).\\
    -0.5+\epsilon_x, & \text{when } 50<i.
    \end{cases}
\end{align*}
A simulation of this setting is visualized in Figure \ref{Figure: Toy Example}. In this simulation $CV(A)=1.47$ and $CV(B)=1.04,$ while $CV_c(A)=2.47$ and $CV_c(B)=3.04.$ Therefore, in case $\bbold^*\perp\bbold,$ while CV selects model B, $CV_c$ selects model A. \footnote{For the setting $\bbold^*=\bbold,$ which will be discussed next, $CV_c=CV$ and both select model B.}
\begin{figure}
\begin{center}
\centerline{\includegraphics[width=0.75\linewidth]{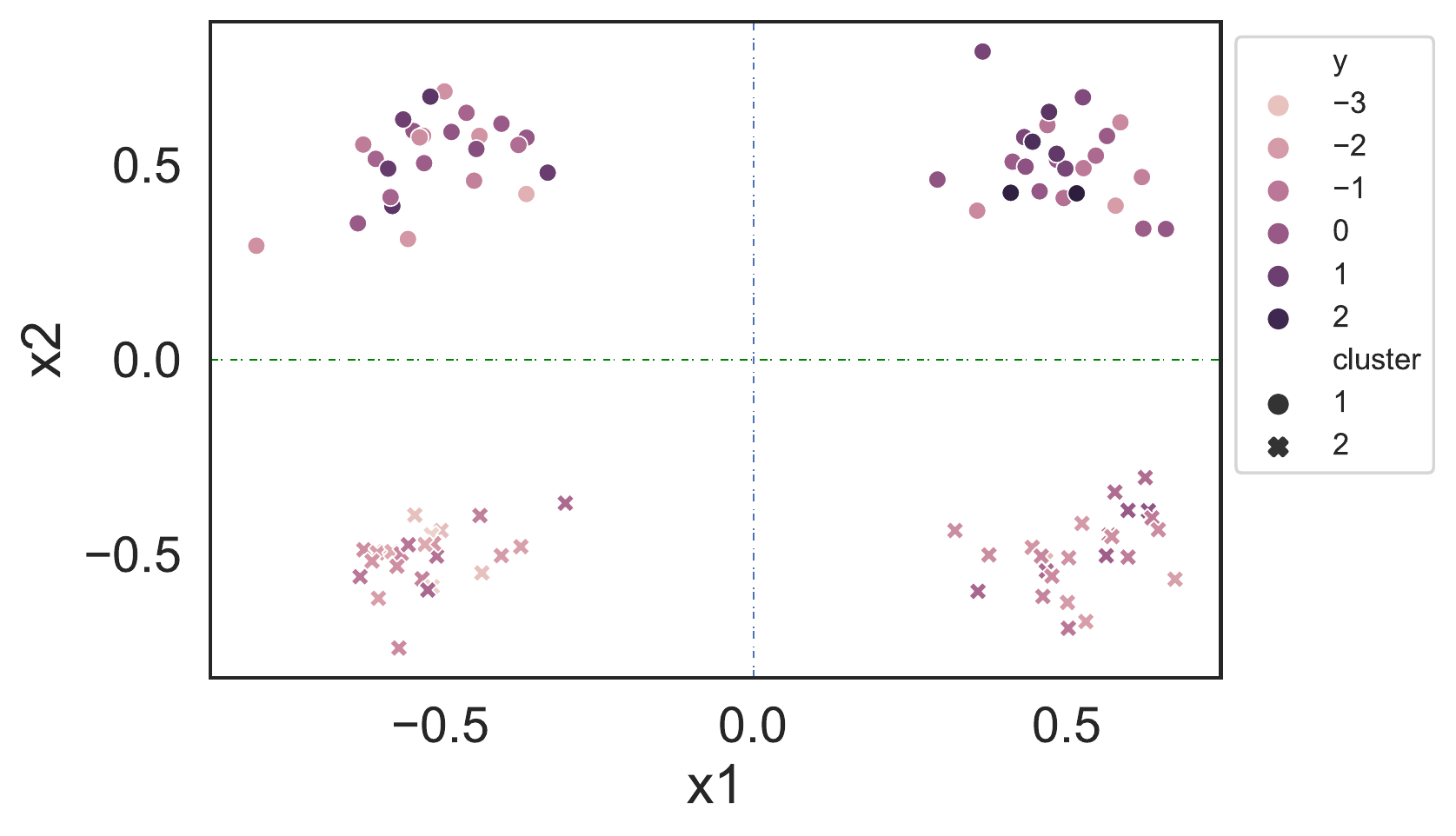}}
\caption{The dashed blue and green lines are the thresholds for model A and model B respectively.}\label{Figure: Toy Example}
\end{center}
\vskip -0.2in
\end{figure}
\end{exmp}

Unlike Cp and $CV_c,$ whose penalties depend on $\Var(\ybold),$ the penalty of AIC in this setting ($\bbold^*\perp\bbold$) is fixed regardless of the training covariance structure at $2p/n.$ Therefore, we can conclude that in AIC the likelihood, $\ell(\ybold;\widehat{\E}(\ybold|X),V),$ is responsible for mixing uncorrelated training set observations in the different paths, while the penalty only affects the stopping rule.  

\subsubsection{\texorpdfstring{$\bbold^*=\bbold$}{TEXT}}\label{Section: The penalty effect, same b}
When $\bbold^*=\bbold$ the correlation between $\ybold^*$ and $\ybold$ is the same as the correlation between observations in $\ybold.$ Therefore, unlike in the $\bbold^*\perp\bbold$ setting, here there is no clear motivation to restrict the tendency to split the nodes based on the correlation structure of the training set (as appears in standard regression trees). Correspondingly, the bias corrections in this setting are also different than in the $\bbold^*\perp\bbold$ setting. For example, $CV$ is not biased in this setting, i.e., $CV_c=CV$ (for more details see Section \ref{Section: Prediction Error Estimation and Model Selection for Correlated Data}). The penalty in Cp, $2\sigma^2\tr(H)/n,$  depends on $\Var(\ybold)$ through $H$ for some models (e.g., for LMM), however for other models it does not depend implicitly on $\Var(\ybold).$ In any case, the effect of $\Var(\ybold)$ is much less prominent than in $\bbold^*\perp\bbold$ setting, where the bias is $2\tr\big(H\Var(\ybold)\big)/n.$ Similarly with the penalty in AIC, $2\tr(H)/n.$ Therefore, in this setting, both Cp and AIC penalties mainly affect to the stopping rule rather than mix between uncorrelated observations. Given additional stopping rules (e.g., tree depth, minimal number of training set observations in each node), we can conclude that the effect of using prediction error estimator instead of training error is limited. A numerical analysis of this scenario is presented in Section \ref{Section: Regression Tree - Same Random Effects}. Still, it is important to emphasize that the proposed model---$f(\xbold^*)+\bbold^{*^{t}}\zbold^*$---is recommended also because of the inference and the use of a solid statistical perspective that the random effects framework enables. 

\subsubsection{\texorpdfstring{$\bbold^*\not\perp\bbold$}{TEXT} but \texorpdfstring{$\bbold^*\neq\bbold$}{TEXT} Scenario}
From a qualitative perspective, this scenario  is the same as the $\bbold^*\perp\bbold$ scenario. In both scenarios, the correlation structure of $\ybold$ is not preserved in the prediction problem. As a result, the bias correction has a key role in balancing the tendency of standard regression trees to split based on the correlation structure of $\ybold.$ The main difference between the scenarios is quantitative and is explicitly expressed in the bias corrections formulas, for example the bias correction in Cp is $2\tr\Big(H\big(\Var(\ybold)-\Cov(\ybold,\ybold^*)\big)\Big)/n.$  This setting of $\bbold^*\not\perp\bbold$ but $\bbold^*\neq\bbold$ is demonstrated in Section \ref{Real Data Results}.

\section{Comparison With Other Algorithms}\label{Section: Comparison to Other Methods}
To our knowledge \citet{sela2012re} were the first to propose integration between LMM and regression tree  by introducing the RE-EM algorithm. The main idea in RE-EM is generating  $\{g_{s}\}_{s=1}^{\mathcal{S}}$ using a standard regression tree that is fitted to the residual, $\ybold-Z\widehat{\bbold}.$ Given $\{g_{s}\}_{s=1}^{\mathcal{S}},$ $\{\mu_s\}_{s\in \mathcal{S}}$ are estimated by GLS. Therefore, the correlation is taken into account in estimating $\{\mu_s\}_{s\in \mathcal{S}},$ but it is ignored in selecting $\{g_{s}\}_{s=1}^{\mathcal{S}}.$ RE-EM algorithm is presented in Appendix \ref{App: Comparison with Other Methods}. \citet{hajjem2014mixed} proposed a RF algorithm which is based on the same logic as RE-EM algorithm.  For numerical comparison between RE-EM and \alname, see Section \ref{section: simulation, Comparison With Previous Algorithms}.

\citet{stephan2015random} proposed the Mixed Random Forest (MRF), that does not ignore the correlation structure when selecting $\{g_{s}\}_{s=1}^{S},$ however still does not address the correlation correctly. The goal in MRF (which is also presented in Appendix \ref{App: Comparison with Other Methods}) is fitting a model that estimates accurately the variance components, rather than optimizing prediction accuracy, as in \alname. Also, MRF assumes a specific data type and  is based on strong distributional assumptions ($\ybold$ is normally distributed and $G=\sigma^2_bI_{q}$). Besides the difference in goals and the assumed settings, the main difference in the tree fitting approach is that MRF uses likelihood loss function for finding $\{g_s,\mu_s\}_{s=1}^S,$ rather than estimated prediction error for correlated data (i.e., training error plus a bias correction) as in \alname. Also, MRF is based on a standard recursive approach, while \alname is based on an iterative approach (for the motivation of using an iterative approach see Section \ref{Section: Trees-Based Linear Mixed Models}). For numerical comparison between MRF and \alname, see Section \ref{section: simulation, Comparison With Previous Algorithms}.

Extensions of these papers,  where the response is binary or count data, as in generalized linear mixed model \citep{wolfinger1993generalized}, where proposed by \cite{fokkema2018detecting,hengl2018random,ngufor2019mixed,speiser2019bimm}.

\section{Numerical Results}\label{Section: Numerical Results}
This section compares the performance of \alname with the standard regression tree algorithm and relevant modifications of it that will be described. The analysis is performed using both simulated data and real data sets for different correlation settings. The simulation part is based on random effects framework and presents results for  $\bbold^*=\bbold$ as well as for $\bbold^*\perp\bbold$ correlation settings. The $\bbold^*\neq\bbold$ but $\bbold^*\not\perp\bbold$ correlation setting is analyzed in the random field context using a real data set with spatial correlation. Also, different prediction error estimator types (Cp, $CV_c$ and AIC) and different trees-based models (regression tree and RF) are analyzed. The code is available in \url{https://github.com/AssafRab/RETCO}.
\subsection{Simulation}\label{Section: simulation}
The training set was generated from the following model:
\begin{align*}
\ybold=&I_{(\xbold_1>0)}+I_{(\xbold_2>0)}+I_{(\xbold_3>0)}+I_{(\xbold_1>0)}I_{(\xbold_2>0)}I_{(\xbold_3>0)}+Z\bbold+\epsilonbold,
\end{align*}
where 
\begin{itemize}
    \item $I_{(\xbold_j>0)},\;\forall j\in [1,2,3]$ is the indicator vector for $\;x_{j,i}>0,\;\forall i\in [1,...,n].$
    \item The sample contains $C$ clusters, each one of size $n_c=n/C.$ $Z\in\mathbb{R}^{n\times C}$ indicates the clusters, i.e., for the first column the first $n_c$ elements are $1,$ and the rest are zero, for the last column the last $n_c$ elements are $1$ and the rest are zero. 
    \item  $\bbold\in\mathbb{R}^{C}$ is the random effects vector, distributed $N_{C}(0,\sigma^2_bI_{C}),$ and
$\epsilonbold\sim N_n(0,I_n).$
\item $\xbold_1,\;\xbold_2$ and $\xbold_3$ are $Z\boldsymbol{\gamma}+\boldsymbol{\eta},$ where $\boldsymbol{\gamma}\sim N_{C}(0,\sigma^2_bI_{C})$ and $\eta_i$ are uncorrelated and  distributed uniformly, $U(-1,1),\forall i\in[1,...,n].$ 
\end{itemize}
 
\subsubsection{New Random Effects (\texorpdfstring{$\bbold^*\perp\bbold$}{TEXT})}\label{Section: simulation, new random effects}
As was mentioned in Section \ref{Section: Linear Mixed Models}, when $\bbold^*\perp\bbold,$ i.e., when $\Cov(\ybold^*,\ybold)=0,$ GLS estimator should be used instead of LMM. Here, Cp prediction error estimator is analyzed and therefore $\ybold^*$ should be related to the same covariate values as in the training set, $\Phi=\{\xbold_1,\xbold_2,\xbold_3,Z\}.$ In order to reduce the variance of the prediction error estimate, the test sample contains $300$ replicates of $\Phi.$ \alname is compared to a standard regression tree with squared error loss function, which is the same loss as in Cp but without the bias correction term. In both algorithms, the stopping rules are depth of tree smaller than $4,$ and number of observations in the terminal node greater than $2.$ The relative difference between the \alname test error and its alternative:
\[
\text{error difference[\%]}=\frac{\text{error(\alname)}-\text{error(standard tree)}}{\text{error(standard tree)}},
\]

is calculated repeatedly for $100$ simulation runs. The average simulation run time is $82.1$ seconds, where \alname takes on the order of 3-8 fold longer to run due to its iterative approach. For more details about \alname's computational complexity, see Section \ref{section: RETCO algorithm}.

\begin{figure}[h]
     \begin{minipage}[l]{0.27\columnwidth}
         \centering
     \includegraphics[width=5cm]{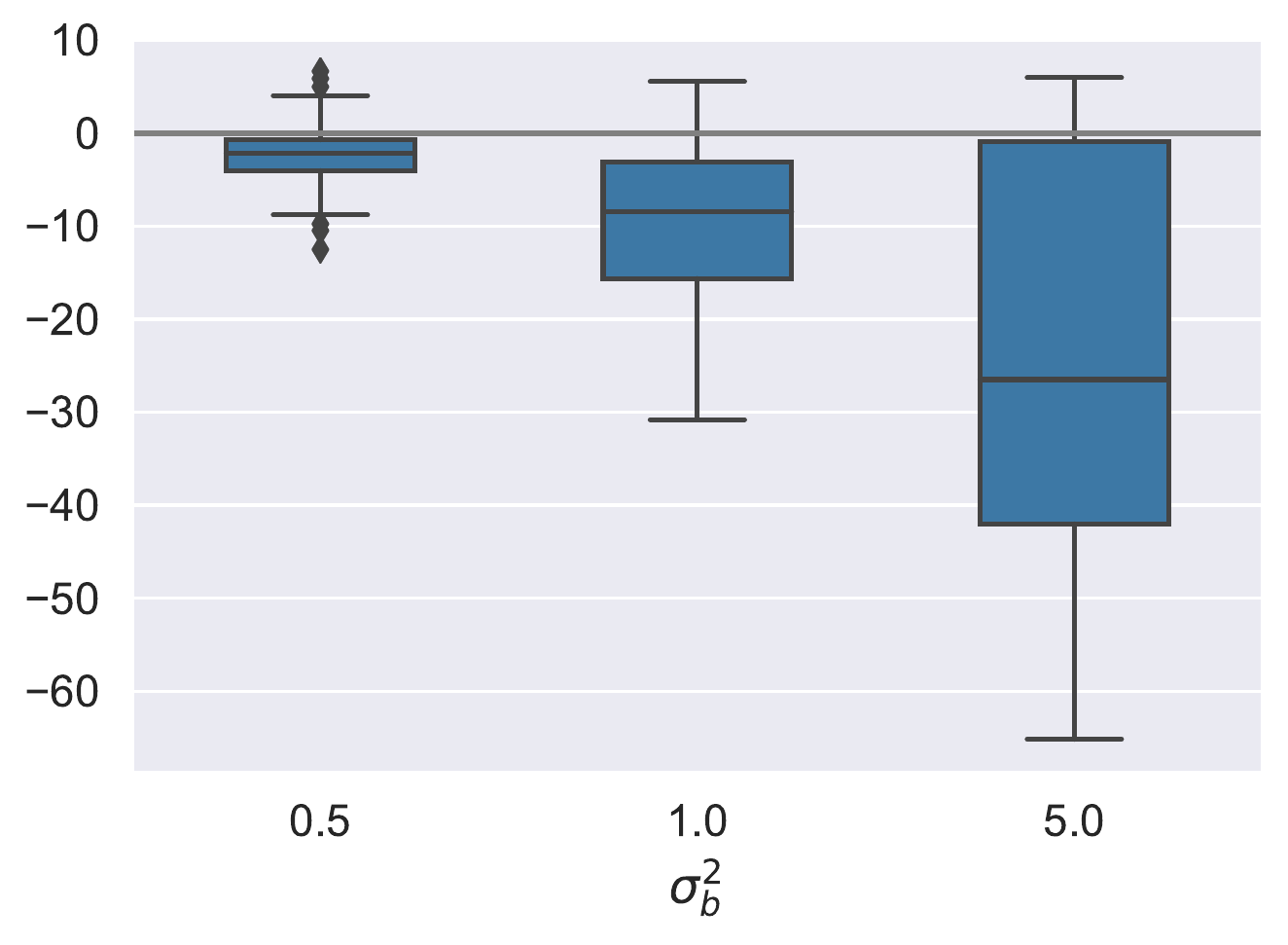}
     \end{minipage}
      \hspace{2em}
     \begin{minipage}[c]{0.27\columnwidth}
         \centering
 \includegraphics[width=5cm]{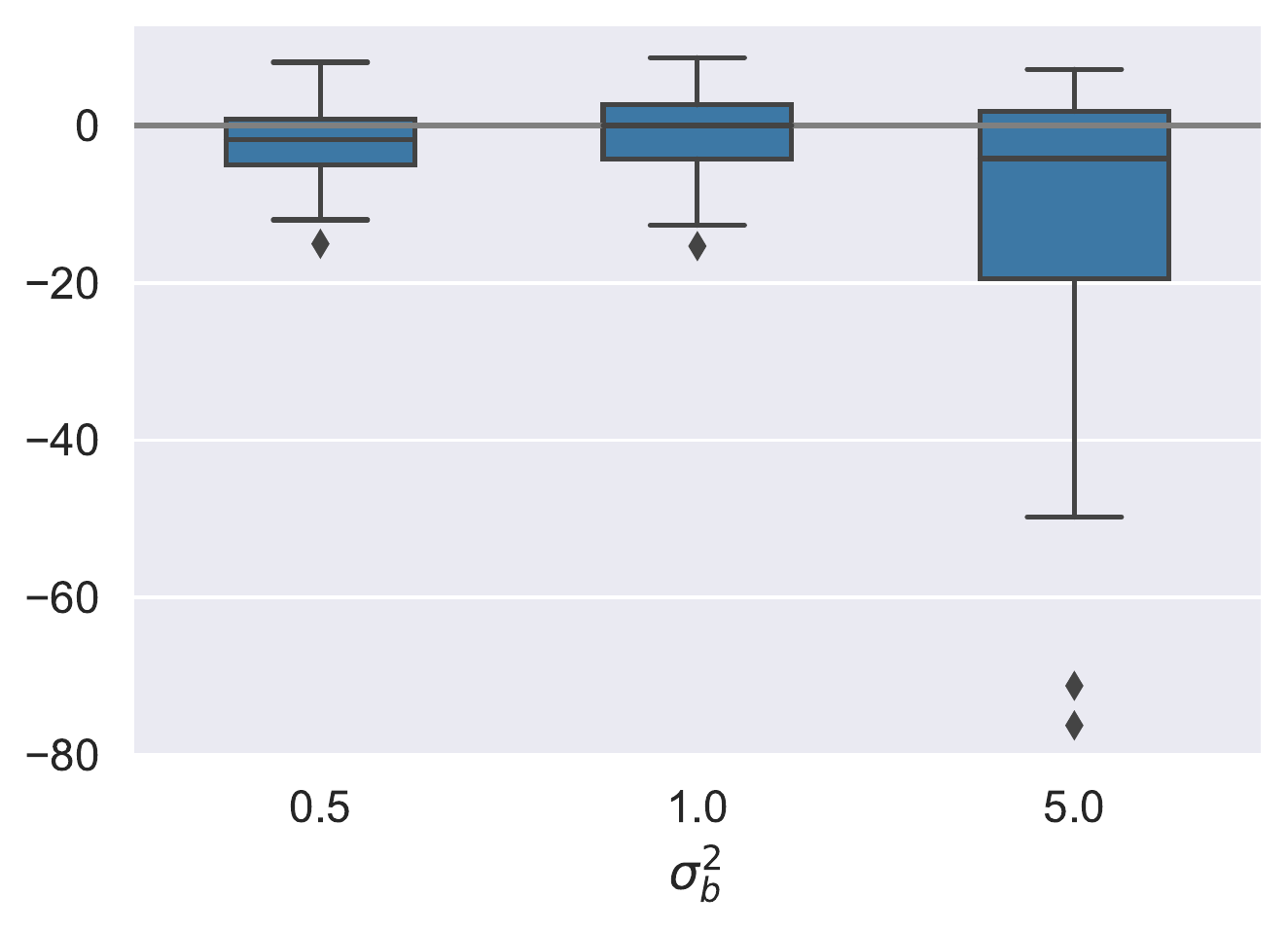}
     \end{minipage}
      \hspace{2.5em}
      \begin{minipage}[r]{0.27\columnwidth}
         \centering
     \includegraphics[width=5cm]{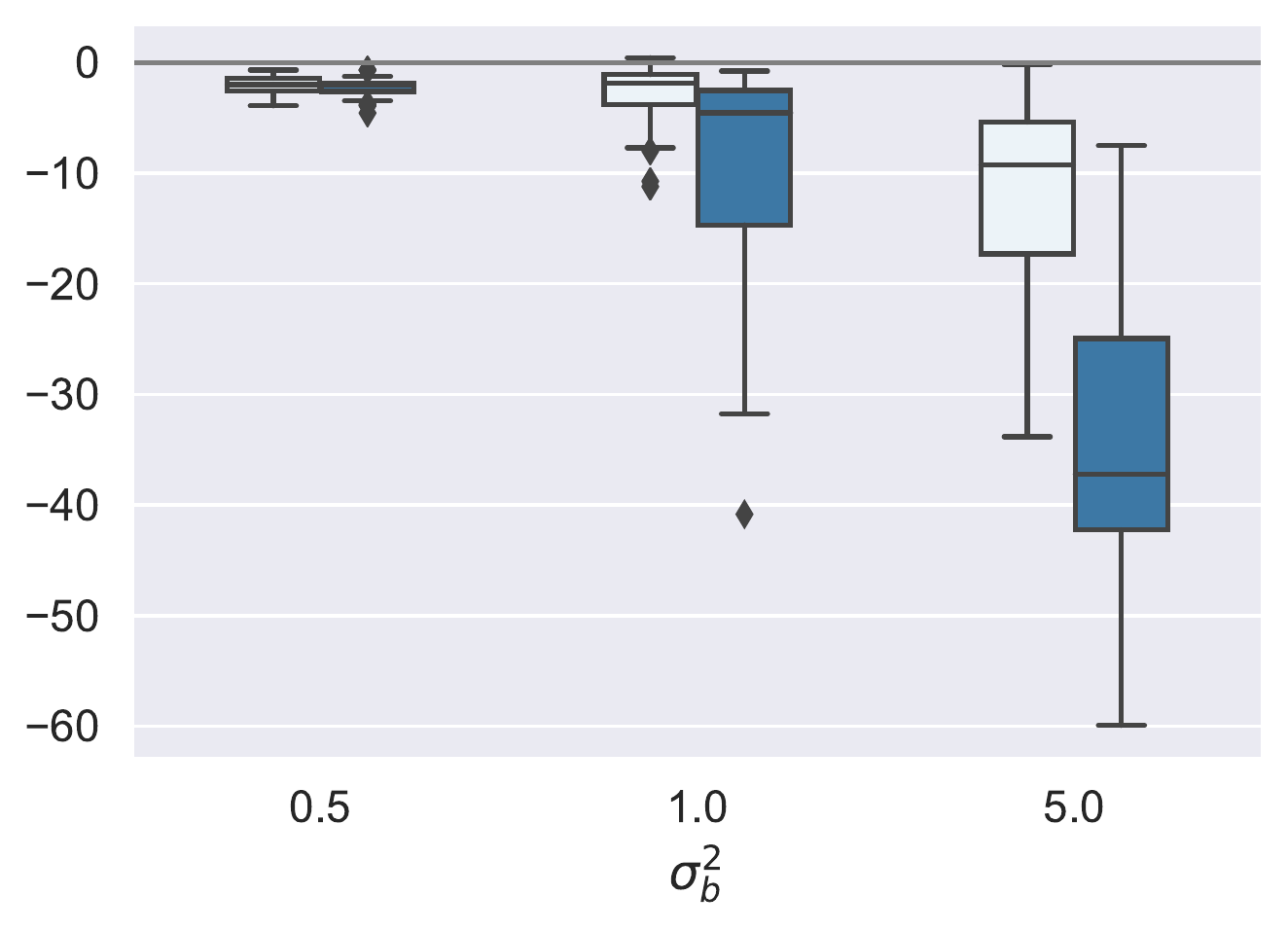}
     \end{minipage}
    \caption{Boxplots of the error difference[\%] for $n=300,\;n_c=50$ and different $\sigma^2_b$ $(0.5,\;1,\;5)$ for different settings. Left: $\bbold^*\perp\bbold$ setting. The means are $-2.5,\;-9.9$ and $-25.2$. Middle: $\bbold^*=\bbold$ setting. The means are $-2.1,\;-1.1$ and $-10.4.$  Right: RF setting. Two versions of \alname are analyzed, in light the version that does not enforce the stopping rule constraint, and in bold the version that enforces the constraint. The means are $-2.1,\;-2.9$ $-11.7$ and $-2.2,\;-9.6$ and $-34.7,$ respectively.} \label{Figure3}
\end{figure}

Figure \ref{Figure3}, left panel, presents boxplots of error difference[\%] for different $\sigma_b^2.$ As expected, when $\sigma_b^2$ is larger (i.e., the correlation is stronger), the improvement in using \alname over the standard regression tree algorithm is bigger. However, also for relatively small $\sigma^2_b$ values \alname outperforms the standard regression tree algorithm. Additional comparisons for different sample sizes ($n$) and cluster sizes ($n_c$), and similar analyses for generalization error setting using $CV_c,$ are given in Appendix \ref{App: Numerical Results}. 

\begin{figure}
\begin{center}
\centerline{\includegraphics[width=1\linewidth]{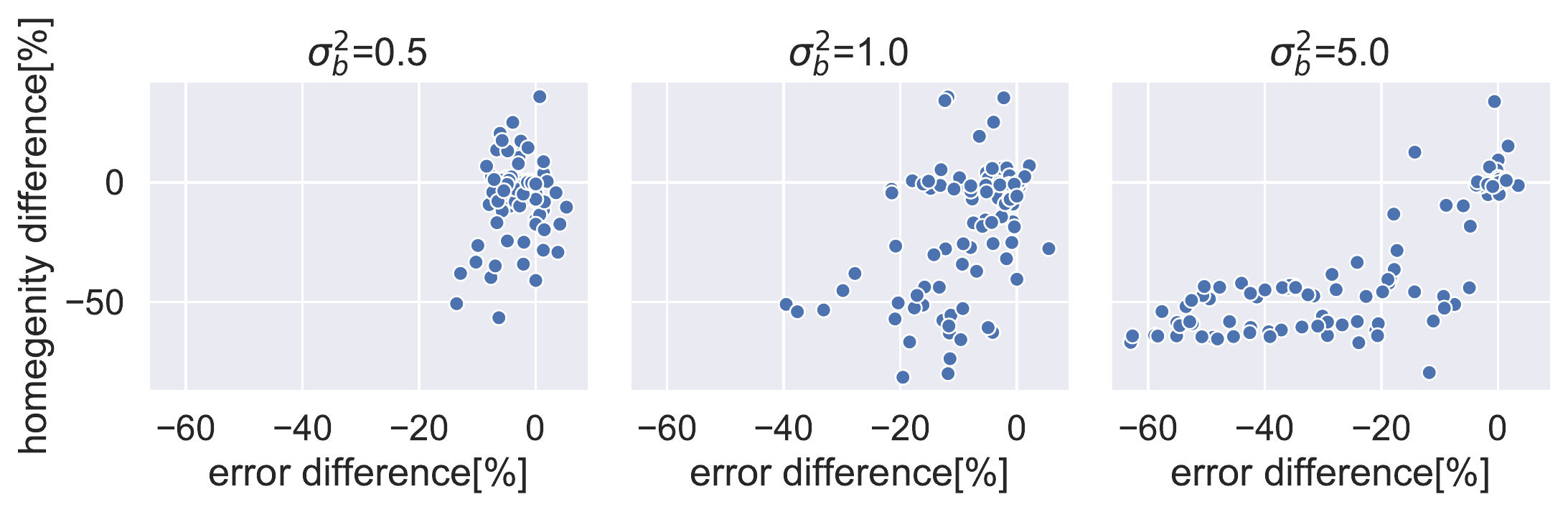}}
\caption{Homogeneity difference[\%] versus error difference[\%] for $n=300,\;n_c=50$ and different $\sigma_b^2$ values.}\label{Figure: Mixing scatter plot}
\end{center}
\end{figure}

As was mentioned in Section \ref{Section: The Correlation Penalty Effect}, \alname balances the tendency of standard tree to split based on the correlation structure of the training sample. Therefore, we expect that \alname mixes training set observations from different clusters in the leaves more than the standard regression tree. The following measure quantifies this mixing property:
\[\small\text{homogeneity}=\sum_{s\in\mathcal{S}}\sum_{c=1}^C|n(c,s)-n(s)/C|,\]
where $n(c,s)$ is the number of training set observations in leaf $s$ that belong to cluster $c$ and $n(s)=\sum_{c=1}^C n(c,s).$ Smaller homogeneity means bigger mixing. Figure \ref{Figure: Mixing scatter plot} plots the training sample homogeneity difference[\%],
\[\small\frac{\text{homogeneity(\alname)}-\text{homogeneity(standard tree)}}{\text{homogeneity(standard tree)}},\] 
versus the error difference[\%] for different $\sigma^2_b$ values.
As can be seen in Figure \ref{Figure: Mixing scatter plot}, error difference[\%] has a positive correlation with homogeneity difference[\%], i.e., the
property of \alname to balance the tendency of standard regression tree to split based on the correlation structure of the training sample is essential.
\subsubsection{Same Random Effects (\texorpdfstring{$\bbold^*=\bbold$}{TEXT})}\label{Section: Regression Tree - Same Random Effects}
For the $\bbold^*=\bbold$ scenario, the training set is the same as in Section \ref{Section: simulation, new random effects}, but the prediction set is different, such that the random effects realizations from the training set are also used for constructing $\ybold^*.$

For \alname, AIC loss function with LMM predictor for $\ybold$ is used for splitting. As presented in Section \ref{Section: Trees-Based Linear Mixed Models}, the predictor in the tree's leaves is GLS and the random effects term is added after fitting the tree. For the standard regression tree, normal likelihood loss function is used with no distinction between random and fixed effects, i.e., all the covariates, including the cluster, can be selected for splitting. 
Correspondingly, the minus log likelihood of i.i.d normal distribution (which is effectively the same as squared error loss) is used as a loss function for the standard regression tree algorithm. The middle panel of Figure \ref{Figure3} compares between the algorithms for different $\sigma_b^2$ values (when $n=300$ and $n_c=50$). 
As we can seen, \alname outperforms the standard tree algorithm. For $\sigma_b^2=0.5$ and $\sigma_b^2=1,$ the average error difference[\%] is relatively small. As was mentioned in Section \ref{Section: The penalty effect, same b}, this phenomenon is expected.
\subsubsection{RF - New Random Effects}
RF is analyzed for $\bbold^*\perp\bbold$ and Cp loss function setting. The training sample model is:
\begin{align*}
\ybold=&I_{(\xbold_1>0)}+I_{(\xbold_2>0)}+I_{(\xbold_3>0)}
+I_{(\xbold_1>0)}I_{(\xbold_2>0)} I_{(\xbold_3>0)}+I_{(\xbold_4>0)} I_{(\xbold_5>0)} I_{(\xbold_6>0)}+Z\bbold+\epsilonbold,
\end{align*}
where $\xbold_i\;\forall i\in\{1,...,6\},\;Z\bbold$ and $\epsilonbold$ have the same distribution as in Section \ref{Section: simulation, new random effects}. Additional parameters that are relevant for RF are:
\begin{itemize}
    \item the maximal tree depth is $10$
    \item The number of regression trees is $T=100$
    \item A random half-sample method is used for sampling the training set for each tree (i.e., the training sample size for each tree is 250 without duplicates)
    \item Three covariates are randomly selected at each split, following the rule of thumb of selecting randomly $round\big(\log_2(p)\big)$ potential covariates at each split. 
\end{itemize}
Also, two versions of \alname are tested: the first uses the stopping rule constraint as presented in \alname inequality (\ref{algo: stopping rule}), the second does not enforce the constraint, and therefore results in deeper trees. The prediction set contains new random effects realizations, such that $\Cov(\ybold^*,\ybold)=0.$ The covariates of the prediction set are $200$ replicates of the covariates of the training set. The analysis was repeated $50$ times. The right panel in Figure \ref{Figure3}
presents boxplots of the error difference[\%] for different $\sigma^2_b.$
As we can seen, both versions of \alname outperform the standard algorithm. Also, forcing the stopping rule gives better results.
\subsubsection{Comparison With Previous Algorithms} \label{section: simulation, Comparison With Previous Algorithms}
The competitors in the left and the right graphs in Figure \ref{Figure3} preserve the main characteristics of the MRF algorithm: taking into account the correlations structure by using GLS estimator and differentiation between random and fixed effects. The exact MRF algorithm, which was designed for a genetic application, is not implemented here since some of its technical details are specific for genetic applications, which are not our main use case. A comparison between \alname and RE-EM algorithm is presented in Appendix \ref{App: Numerical Results}. As expected, \alname's performance is uniformly superior to both algorithms due to its use of prediction error estimates for splitting and the careful consideration of correlation structures in splitting and prediction. 

\subsection{Real Data Analysis}\label{Real Data Results}
This section presents real data analyses comparing the performance of the standard regression tree and RF algorithms to their \alname versions for six different data sets with various correlation structures. The data sets and the prediction problems are briefly described in Section \ref{Datasets Description}, additional technical information can be found in Appendix \ref{App: Numerical Results}. 

Table \ref{Table: real data resutls} summarizes the results. As can be seen, for all the six analyses the test errors of the standard regression tree and RF algorithms are greater than the test errors of their \alname versions (negative error difference [\%]), moreover in several analyses the improvement of \alname over the standard algorithm is very large.

\begin{table}
\begin{center}
\begin{small}
\begin{tabular}{|llcc|}\toprule
 Correlation Structure &  Data Set Name& Regression Tree &RF\\\midrule
\multirow{2}{*}{Clusters}& FIFA& $-4.9\%$ &$-7.1\%$\\
& Crimes &$-8.1\%$ &$-4.4\%$\\\hline 
\multirow{2}{*}{Spatial}& Korea Temperature&$-14.5\%$&$-13.6\%$\\
& California Housing &$-14.2\%$&$-3.3\%$\\\hline
\multirow{2}{*}{Longitudinal} &   Parkinson's Disease&$-32.9\%$&$-35.4\%$\\
&Wages & $-20.1\%$&$-7.9\%$\\\bottomrule    \end{tabular}
\end{small}
\end{center}
\caption{Error difference [\%] between \alname and standard regression tree, and between \alname and standard RF.}\label{Table: real data resutls}
\end{table}
\subsubsection{Prediction Problems Description}\label{Datasets Description}
\begin{itemize}
\item \textit{FIFA} --
This data set contains football players' market-values. The data set has a clustered correlation structure, where the cluster variable is the player's club, such that market-values of players from the same club are correlated but from different clubs are not correlated. The prediction goal is to predict the market-values of new players from new clubs. In order to satisfy this prediction goal, the training set contains the observations of players from $20$ clubs that were randomly sampled ($548$ observations), and the test set contains the observations of the other clubs ($17,939$ observations). Since the covariate values of the prediction set are not the same as the covariate values of the training set, CV-type loss function is used in the algorithms' splitting criterion -- $CV_c$ for \alname and CV for the standard regression tree/RF algorithms. The prediction error is estimated by the average squared errors of the test set. The data set is publicly available on \href{https://www.kaggle.com/}{Kaggle}.
\item \textit{Communities and Crime} -- This data set presents the number of violent crimes per population size in US communities. The data set has a clustered correlation structure, where the clusters are the US states (each state contains many communities). The training set contains $15$ clusters that were randomly sampled ($790$ observations), where the test set contains the other clusters ($1,204$ observations). For the same reason as in the FIFA data set, CV-type loss function is used in the algorithms' splitting criterion. The data set was introduced by \cite{redmond2002data}, and is publicly available on the \href{https://archive.ics.uci.edu/ml/index.php}{UCI repository}.
\item \textit{South Korea Temperature ('bias correction of numerical prediction model temperature forecast')} --  This data set contains daily maximal temperature measurements (collected in August between the years $2013-2017$) at several sites in South Korea. The prediction goal is to predict the maximal temperature of new days. Measurements of the first two years were selected ($575$ observations) in order to predict the maximal temperature of the same set of days in the next years ($2,325$ observations). Due to the spatial correlation structure, exponential kernel covariance function was used for modeling. Since all the records are measured at the same sites, in-sample error type is used in the algorithms' splitting criterion ($Cp$ for \alname and squared error loss for the standard regression tree/RF). The data set was introduced by \cite{cho2020comparative}, and is publicly available on \href{https://archive.ics.uci.edu/ml/index.php}{UCI repository}.

\item \textit{California Housing Prices} --
This data set contains the median house value within a block for different blocks in California. Some of the blocks belong to the same clusters (same coordinate values), therefore the data set has a clustered-spatial correlation structure, which can be represented by the following kernel covariance function:
\begin{align*}
\Cov(y_i,y_j)=&\mathcal{K}(|\zbold_i-\zbold_j|)+\sigma_b^2I_{cluster(i)=cluster(j)}+\sigma_{\epsilon}^2I_{i=j},
\end{align*}
where $\mathcal{K}(\cdot)$ is the exponential kernel covariance function, $\zbold_i,\zbold_j$ are the coordinates and $cluster(i),$ $cluster(j)$ are the clusters of $y_i,y_j.$ The prediction goal is to predict the median house value of new blocks from new clusters. Therefore, $100$
clusters are randomly sampled ($279$ observations) for the training set and the other clusters are used as the test set ($12,124$ observations). Since the prediction goal is to predict median house values from new clusters, then $\bbold^*\neq\bbold.$ However, due to the spatial correlation, the prediction set median house values are correlated with the training set median house values, and therefore this setting satisfies the  $\bbold^*\neq\bbold\cap\bbold^*\not\perp\bbold$ scenario (see Section \ref{new random effects}). CV-type loss function is used in the algorithms' splitting criterion since the covariate points of the training set and the prediction set are different. The data set was introduced by \cite{pace1997sparse},
and is publicly available on \href{https://www.kaggle.com/}{Kaggle}.
\item \textit{Parkinson's Disease Telemonitoring} -- This longitudinal data set contains Parkinson's disease symptom scores of $42$ individuals along six-months trial. The clustered-temporal correlation structure, where the clusters refer to the individuals, can be modeled by LMM with random intercept for the cluster and random slope for the time variable. The prediction goal is predicting the score of new individuals, therefore five individuals were randomly sampled ($742$ observations) for the training set and the others were designated as the test set ($5,179$ observations). The covariates in this data set are biomedical voice measurements and their values are approximately the same for all the individuals. Therefore, $Cp$ and  squared error loss loss functions are used in the splitting criterion for \alname and standard regression tree/RF, respectively. The data set was introduced by \cite{tsanas2009accurate}, and is publicly available on the \href{https://archive.ics.uci.edu/ml/index.php}{UCI repository}.
\item \textit{Wages} -- This longitudinal data set presents the average hourly wages by year of $888$ employees. As in the Parkinson's Disease Telemonitoring data set, this data set can be modeled by LMM with random intercept and random slope, where the employee is the cluster variable. The prediction goal is to predict the average hourly wage of new employees. $50$ individuals were randomly sampled ($331$ observations) for the training set and the other are used as the test set ($6,071$ observations). Since the covariate values of the training set and the prediction set are different, CV-type loss function is used in the algorithms' splitting criterion. The data set was introduced by \cite{singer2003applied}, and is publicly available in \href{https://cran.r-project.org/web/packages/brolgar/index.html}{brolgar package in R software}.
\end{itemize}

\section{Conclusions}
This paper presents a new algorithm, \alname, for fitting regression trees-based models for correlated data. Analyzing various settings with different correlation structures lead to the conclusion that \alname substantially improves prediction performance in settings involving correlated data. 

Unlike standard regression trees-based models, which ignore the correlation structure of the data, \alname accounts for the correlation structure in various ways, such as using prediction error estimates for correlated data as the loss function in the  splitting criterion. As discussed and demonstrated, using prediction error estimators for correlated data instead of training error neutralizes the tendency to fit a tree that divides the training set based on its correlation structure, as is likely to happen in standard regression trees-based models. 

Extensive data analysis, including analysis of six different real data sets, shows the superiority of \alname over standard regression trees-based model, as well as its generality that enables to implement it under various settings.

\acks{This work was supported by the Israel Science Foundation, grant 1804/16 and by the European Union Seventh Framework
Programme grant agreement no. 785907 (Human Brain Project).}


\newpage

\appendix
\section{Theoretical Background}\label{App: Theoretical Background}
This appendix extends the theoretical background that is given in Section \ref{Section: Theoretical Background}.
\subsection{Regression Tree, Random Forest and Gradient Boosting}
Algorithm \ref{Algo: Standard tree} presents a typical regression tree algorithm. 
\begin{algorithm}[h!]
\caption{A typical tree-based algorithm}
\label{Algo: Standard tree}
\begin{algorithmic}
\STATE {\bfseries Input:} $\ybold,\;X.$
\STATE {\bfseries Output:} $f(\cdot).$ 
\STATE {\bfseries High-level setting:} select a training error loss function and define stopping rules
\STATE {\bfseries Initialization:}
 $\mathcal{S}=\{g_1,\mu_1\},$ where $g_1=\mathbb{R}^p,\;\mu_1=\sum_{i=1}^n y_i/n.$
 \REPEAT
 \STATE  
 \begin{enumerate}
     \item Given the predefined stopping rules, for each node $s\in \mathcal{S}$ solve the following optimization problem: 
 \begin{align*}
 c_s,j_s=\underset{c\in\mathbb{R},j\in J_s}{\argmin}\frac{1}{|I_s|}\sum_{i\in I_s}Loss\big(y_i,I_{(x_{i,j}\leq c)}\mu_s^l(c)+I_{(x_{i,j}>c)}\mu_s^r(c)\big),
 \end{align*}
where $J_s$ is the set of available covariates for splitting node $s,$ $I_{s}=\{i|\xbold_{i}\in g_{s}\},$ $\mu_s^l(c)$ and $\mu_s^r(c)$ are the mean estimators of $\{y_i|\xbold_i\in g_s \cap x_{i,j}\leq c\}$ and $\{y_i|\xbold_i\in g_s \cap x_{i,j}> c\}$ respectively.
\item  Update $\mathcal{S}$ by replacing $(g_{s},\mu_{s})$ by the new two nodes: $\big(g_{s}\cap x_{j_{s}}\leq c_{s},\mu_{s}^r(c_{s})\big),$ $\big(g_{s}\cap x_{j_{s}}> c_{s},\mu_{s}^l(c_{s})\big),$ where $x_{j_{s}}$ is the covariate $j_{s}$ and $\mu_{s}^r(c_{s}),\mu_{s}^l(c_{s})$ are the related mean predictors.
 \end{enumerate}
 \UNTIL {Stopping rules are satisfied $\forall s\in\mathcal{S}$}
\end{algorithmic}
\end{algorithm}

Random forest (RF) and gradient boosting (GB) predictor are based on averaging an ensemble of trees:
\[
f(\xbold^*)=\sum_{t=1}^T\lambda_t\sum_{s=1}^{S_t} I_{(\xbold^*\in g_{t,s})}\mu_{t,s},
\]
where $T$ is the number of trees and $\lambda_t\in (0,1]$ is the learning rate (for RF $\lambda_t=1/T,\;\forall t$). 

The regression trees in RF and GB are fitted in different ways than in a standard regression tree. In RF, the training set of each tree is sampled from the original sample (e.g., sampling with replacement of size n, half-sample), and the set of the potential covariates of each split is a random sample of $J_s.$  In GB the trees are dependent and created consecutively, where the dependent variable of each tree is the residual of the previous tree. Also, there are many techniques for reducing over-fitting and model variance which are relevant for RF and GB, but not relevant for standard regression tree model. For more information about RF and GB see \citet{freund1999short,friedman2001greedy,breiman2001random,Hastie2009elements}. 

Note, unlike in regression tree model, which tends to over-fit and therefore suffers from high variance, RF has relatively low variance due to the averaging over the $T$ trees. This property affects the optimal structure of trees in RF. While the tree depth in regression tree model should be restricted in order to avoid over-fitting, the trees in RF can be large whenever $T$ is respectively large \citep{criminisi2011decision}. Since the trees in RF are correlated, the RF variance decreases in a smaller rate than $T.$ Commonly the trees' depth in RF is also restricted for various reasons, such as computational cost that RF with deep trees (and consequently large $T$) requires.

\subsection{Prediction Error Estimation and Model Selection for Correlated Data}
\subsubsection{Cp}
The original Cp, when $\Cov(\ybold,\ybold)=\sigma^2\times I_n$ and $\Cov(\ybold^*,\ybold)=0,$ was introduced by \citet{mallows1973some} is:
\begin{align*}
Cp=\frac{1}{n}\|\ybold-\widehat{\ybold}\|_2^2+\frac{2\sigma^2}{n}p.
\end{align*}

\subsubsection{AIC}
The standard AIC under normality and i.i.d assumptions, that was introduced by \citet{akaike1974new} is:
\[
AIC=-\frac{2\ell\big(\ybold;\widehat{\E}(\ybold|X),\sigma^2\big)}{n}+\frac{2p}{n},
\]
where $\ell\big(\ybold;\widehat{\E}(\ybold|X),\sigma^2\big)$ is the log-likelihood of $\ybold.$



\section{Estimating Variance Components for Clustered Data}
\label{App: Estimating variance components for the random intercept model}

When $\ybold$ has  a clustered correlation structure, i.e., its covariance matrix follows:
\[
\Cov[i,j]=\begin{cases}
    \sigma^2_{\epsilon}+\sigma^2_{b}, & \text{when } i=j\\
    \sigma^2_{b}, & \text{when } i\neq j \text{ but } c(i)=c(j) \\
    0,& \text{ o.w }
    \end{cases},
\]
where $c(i)$ is the cluster that observation $i$ belongs to and $\sigma_{\epsilon},\;\sigma_{b}$ are in $R^{+},$
then $\sigma^2_{\epsilon}$ and $\sigma^2_{b}$ can be estimated in  a closed-form way. In order to simplify the equations let us assume $\E\ybold=0.$ In this case: 
\[
\widehat{\sigma}_{\epsilon}^2=\frac{\sum_{i=1}^n\big(y_i-\Bar{y}(i)\big)^2}{(n-C)},
\]
where $\Bar{y}(i)$ is the average of the cluster that $y_i$ belongs to, and $C$ is the number of clusters, 
\[
\widehat{\sigma}_{b}^2=\Big(\frac{\sum_{i=1}^{n}(\Bar{y}\big(i)-\Bar{y}\big)^2}{C-1}-\widehat{\sigma}_{\epsilon}^2\Big)\times\frac{C-1}{n-\sum_{j=1}^{C}n_j^2/n},
\]
where $\Bar{y}$ is the average of $\ybold$ and $n_j$ is the number of observations in cluster $j.$
When $\E(\ybold)\neq0,$ the variance parameters should be calculated for the residual, $\ybold-\widehat{E}{\ybold}.$

\section{Comparison with Other Methods}\label{App: Comparison with Other Methods}
Algorithm \ref{RE-EM} presents  the RE-EM algorithm, which was proposed by 
\citet{sela2012re}. Algorithm \ref{Mixed Random Forest} presents MRF algorithm for a single tree, that was proposed by \citet{stephan2015random}.\footnote{\citet{stephan2015random} do not supply an organized algorithm, Algorithm \ref{Mixed Random Forest} tries to formalize their approach as given in their supplementary material.} For MRF, in order to simplify notations, denote $\ybold,\;X,\;Z$ as the bootstrap sample and ignore the features sampling at each split.

\begin{algorithm}[tb]
\caption{RE-EM Algorithm}
\label{RE-EM}
\begin{algorithmic}
\STATE {\bfseries Input:} $\ybold,\;X,\;Z.$
\STATE {\bfseries Output:} $f(\cdot),\;\widehat{\bbold}.$
 \STATE {\bfseries Initialization:} $\widehat{\bbold}^{(0)}=0.$
  \REPEAT
  \STATE  \begin{enumerate}
      \item  Using the fixed effects covariates, fit CART algorithm \citep{breiman1984classification} for $(y-Z\widehat{\bbold}^{(k-1)})$ and extract $\{g^{(k)}_{s}\}_{s=1}^{S^{(k)}}$ from the fitted tree.\label{CART step}
      \item Estimate $\{\mu_{s}\}_{s=1}^S,$ $G$ and $\sigma^{2}$ using the following LMM model:
 \[
  y_{i}=\sum_{s=1}^{S}I_{(\xbold_{i}\in g_{s}^{(k)})}\mu_{s}+\zbold_{i}\bbold+\epsilon_{i},
  \]
  where $\bbold\sim N_q(0,G),\;\epsilon_i\sim N(0,\sigma^2).$
  \item  Given $\{\mu_{s}^{(k)}\}_{s=1}^{S^{(k)}},\;\widehat{G}^{(k)}$ and $\widehat{V}^{(k)}=Z\widehat{G}^{(k)}Z^{t}+\widehat{\sigma}^{2,(k)}I_n:$ estimate $\widehat{\bbold}^{(k)}$ using the BLUP formula.
  \end{enumerate}
 \UNTIL{Convergence of $\widehat{\bbold}^{(k)}.$}
\end{algorithmic}
 \end{algorithm}

\begin{algorithm}[tb]
 \caption{Mixed Random Forest (algorithm for fitting a single tree)}
 \label{Mixed Random Forest}\begin{algorithmic}
\STATE {\bfseries Input:} $\ybold,\;X,\;Z.$
\STATE {\bfseries Output:} $f(\cdot),\;\widehat{\sigma}^2_b,\widehat{\sigma}^2.$
 \STATE {\bfseries Initialization:} $g_1=\{0,1\}^J.$ Also, estimate $\delta=\sigma^2/\sigma^2_b.$
  \REPEAT
 \STATE  
 \begin{enumerate}
     \item  Given the predefined stopping rules, for each node $s\in \mathcal{S},$ find the following parameters: 
 \begin{align*}\small
 \tilde{j_s},\tilde{\sigma}^2=&
 \underset{j_s\in J_s,\sigma^2\in\mathbb{R}^+}{\text{argmax}}\;\ell(\ybold;\sigma^2,\mu_s^l, \mu_s^r,\{\mu_l\}_{l\in \mathcal{S}/s},
g_s\cap(x_{j_s}=0),g_s\cap(x_{j_s}=1),\{g_{l}\}_{l\in \mathcal{S}/s}),
 \end{align*}
 where
$\mu_s^l,\;\mu_s^r$ and $\{\mu_l\}_{l\in \mathcal{S}/s}$ are the GLS estimators for $\{y_i|\xbold_i\in g_s\cap(x_{i,j_s}=0)\},$ $\{y_i|\xbold_i\in g_s\cap(x_{i,j_s}=1)\},$ and $\{y_i|\xbold_i\in g_{l}\}_{l\in \mathcal{S}/s}$ subsets respectively.
\item  Update $f(\cdot):$ replace
     each node $s\in\mathcal{S}$ by $\{g_s\cap(x_{\tilde{j}_s}=0),\mu_s^l\}$ and
 $\{g_s\cap(x_{\tilde{j}_s}=1),\mu_s^r\}.$
 \end{enumerate}
 \UNTIL{Stopping rules are satisfied $\forall s\in\mathcal{S}.$}
\end{algorithmic}
 \end{algorithm}

\section{Numerical Results}\label{App: Numerical Results}
This appendix presents additional results that are related to Section \ref{Section: simulation}, as well as detailed information relating the settings in Section \ref{Real Data Results}.
\subsection{Regression Tree - New Random Effects (\texorpdfstring{$\bbold^*\perp\bbold$}{TEXT})}
\subsubsection{In-Sample Error Setting:}
\begin{figure*}
     \begin{minipage}[l]{0.27\columnwidth}
         \centering
         \includegraphics[width=5cm]{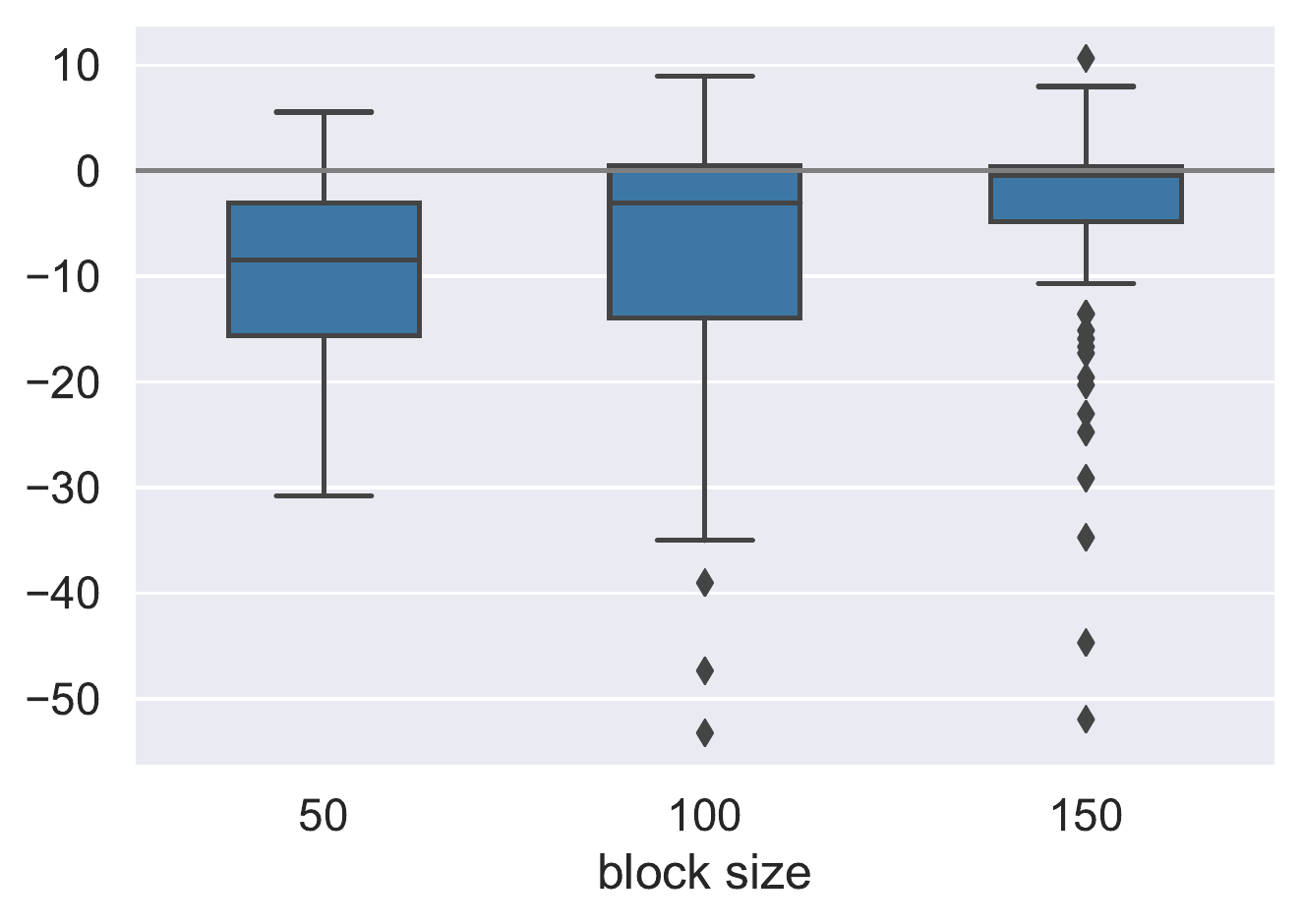}
     \end{minipage}
     \hspace{2em}
     \begin{minipage}[c]{0.27\columnwidth}
         \centering
         \includegraphics[width=5cm]{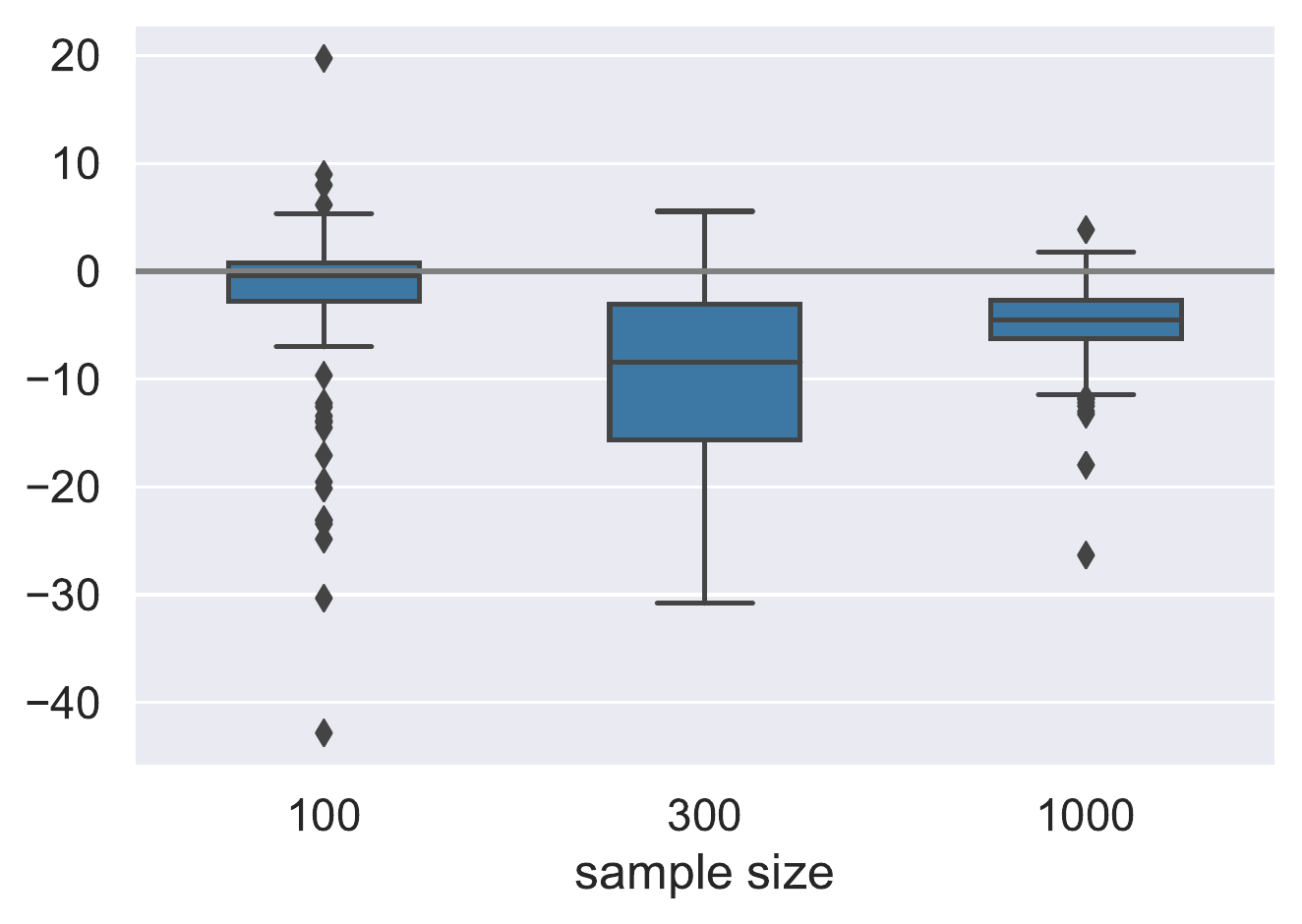}
     \end{minipage}
     \hspace{2.5em}
          \begin{minipage}[r]{0.27\columnwidth}
         \centering
         \includegraphics[width=5cm]{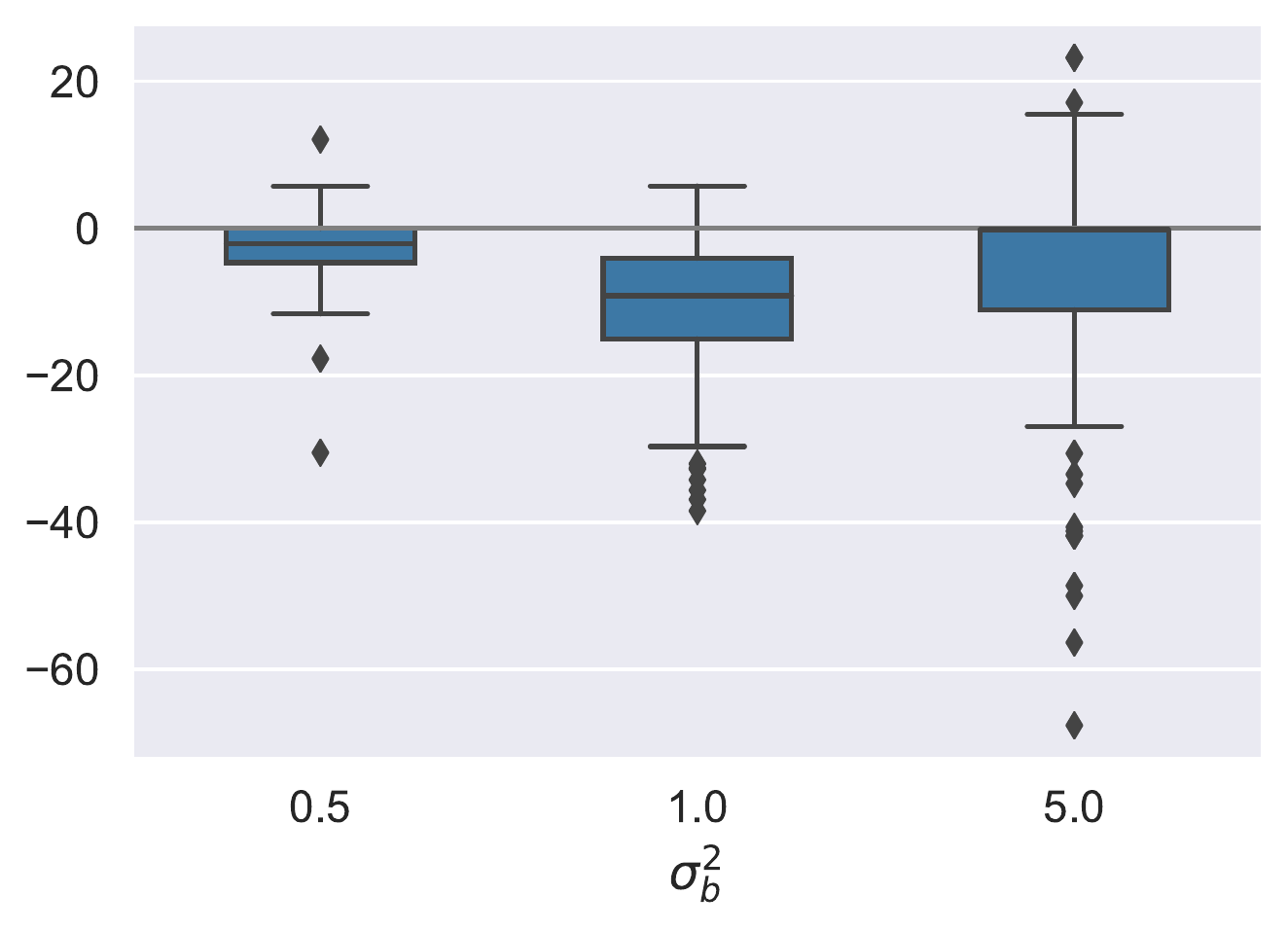}
     \end{minipage}
    \caption{Boxplots of the error difference[\%]. Left: In-sample error setting, $\bbold^*\perp\bbold,$
  $n=300,\;\sigma_b^2=1$ and different $n_c.$ The means are $-9.9,\;-8.0$ and $-4.3.$ Middle: In-sample error setting, $\bbold^*\perp\bbold,$
  $\sigma_b^2=1\;n_c=50$ and different $n.$ The means are $-2.9,\;-9.9$ and $-5.0.$ Right: Generalization error setting, $\bbold^*\perp\bbold,$
  $n=300,\;n_c=50$ and different $\sigma_b^2.$ The means are $-2.8,\;-11.0$ and $-7.3.$} \label{Figure: Sup}
    \vskip-0.1in
\end{figure*}

Given the simulation setting that is described in Section \ref{Section: simulation, new random effects}, Figure  \ref{Figure: Sup}, left panel, presents the effect of the cluster size ($n_c$), on the performance of \alname. As can be seen in the figure, for larger block size ($n_c=150$) the average error difference[\%] is smaller.

Figure \ref{Figure: Sup}, middle panel,
presents the effect of the sample size on the performance of \alname. As can be seen, \alname performs better for all the settings. 
Also, as expected, the error difference[\%] variance is smaller for larger sample sizes. The variance depends on the maximal depth of the tree, which was set to three. Tree with three levels has potentially eight predictors, which is a large amount of predictors when $n=100,$ but small when $n=1000.$ Therefore, when $n=100$ the trees are noisy for both algorithms, and their relative difference is noisy as well.

\subsubsection{Generalization Prediction Error Setting:}
In order to analyze \alname performance in generalization prediction error setting, the test set setting was changed such that the prediction set covariates are nonidentical to the training sample covariates (but are sampled from the same distribution). As was described in the paper, $CV_c$ loss function estimates the generalization error unbiasedly by correcting the standard CV error. Therefore, $CV_c$ loss function is used in \alname and CV loss function is used for the standard regression tree algorithm. Figure \ref{Figure: Sup}, right panel, presents
the error difference[\%] for different $\sigma^2_b.$ 
\subsection{Comparison With RE-EM}
Figure \ref{Figure, RE-EM}, left figure, presents the error difference between \alname and RE-EM for the scenario when $\bbold^*\perp\bbold.$ All the other setting details are the same as in Section \ref{Section: simulation, new random effects}. Similarly, Figure \ref{Figure, RE-EM}, right figure, presents the error difference between \alname and RE-EM for the scenario when $\bbold^*=\bbold.$ All the other setting details are the same as in Section \ref{Section: Regression Tree - Same Random Effects}. As can be seen in Figure \ref{Figure, RE-EM}, RETCO performs better than RE-EM. 
\begin{figure}
     \begin{minipage}[l]{0.5\columnwidth}
         \centering
         \includegraphics[width=5cm]{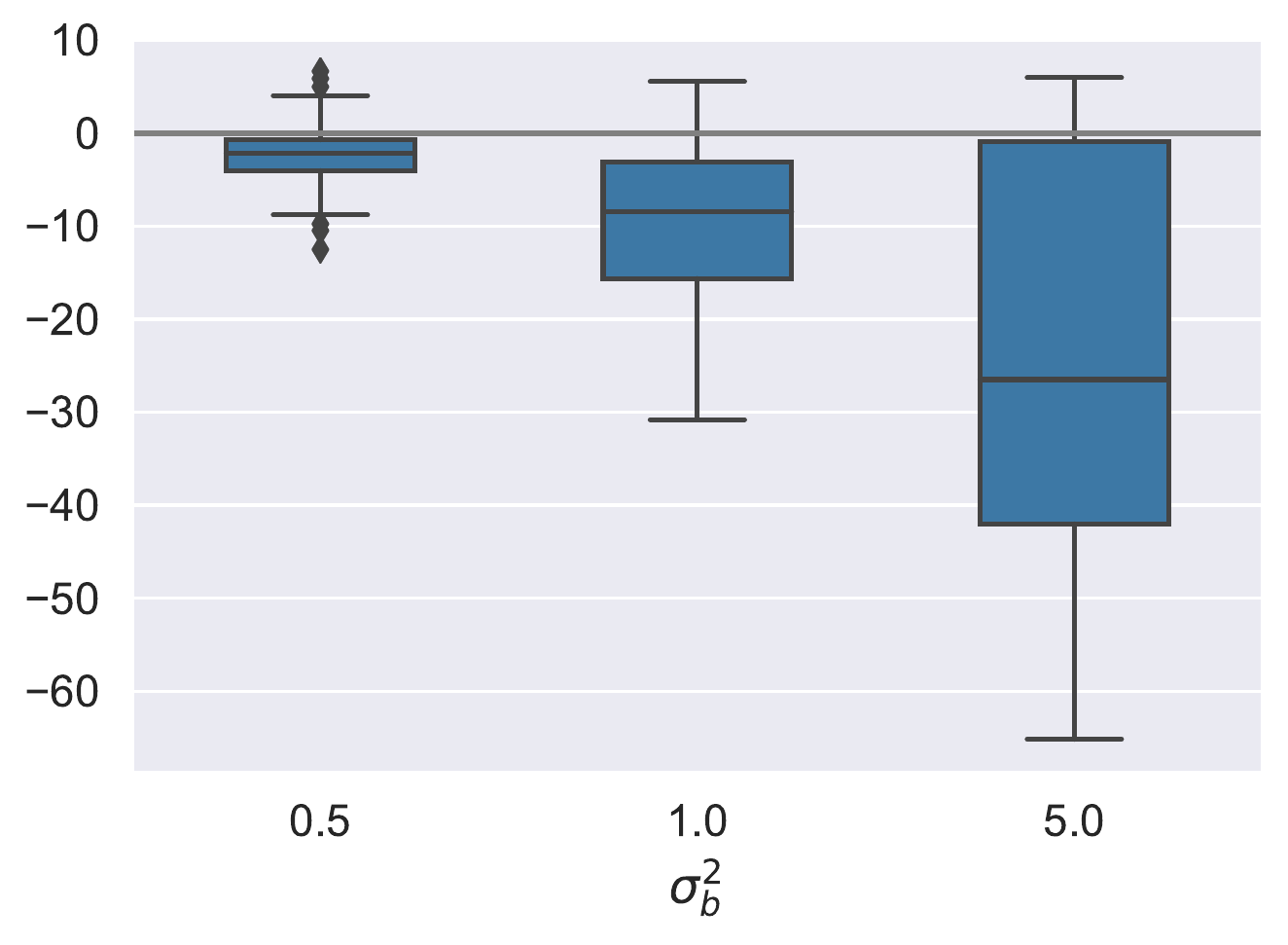}
     \end{minipage}
     \begin{minipage}[c]{0.5\columnwidth}
         \centering
         \includegraphics[width=5cm]{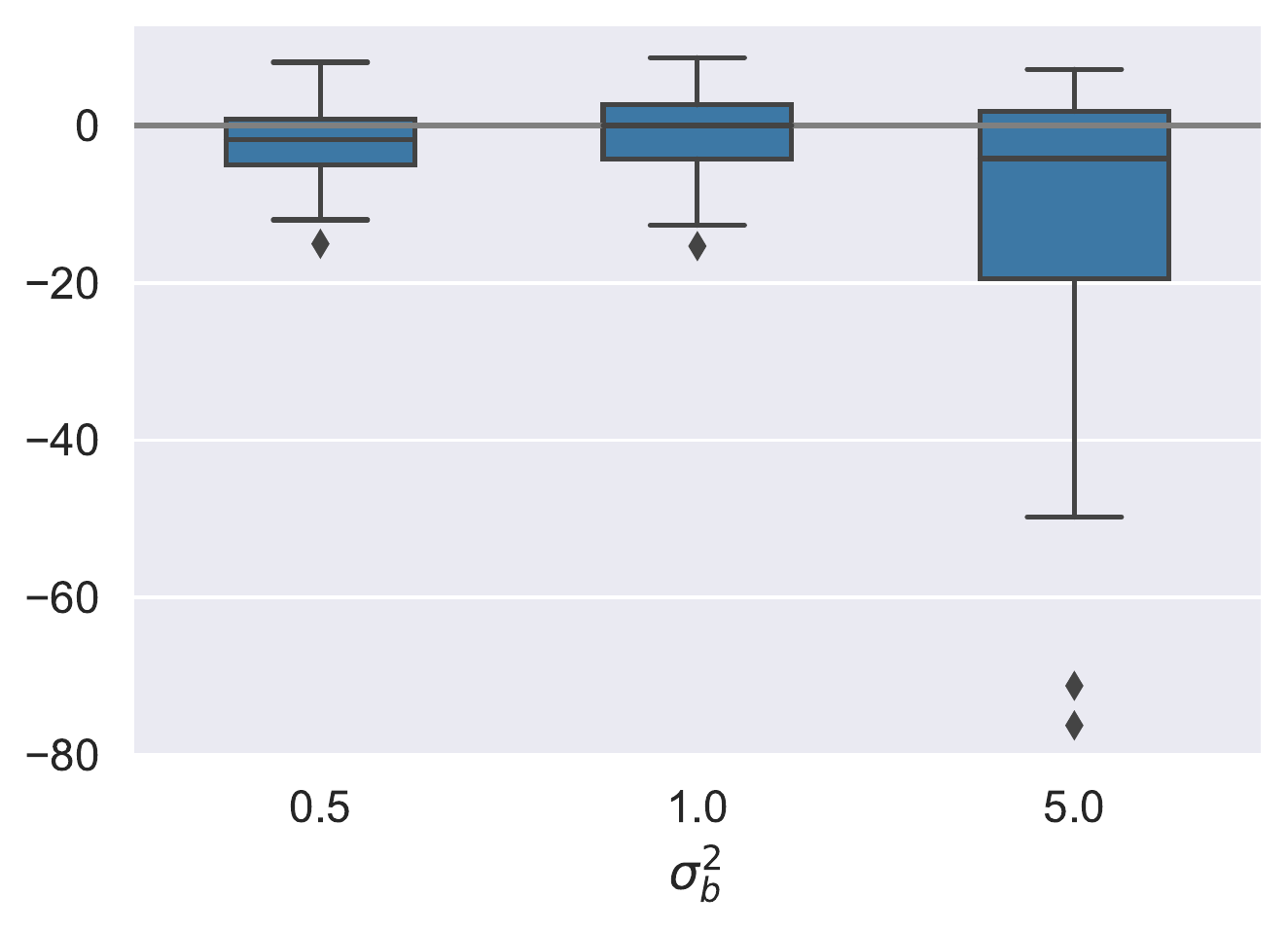}
     \end{minipage}
    \caption{Boxplot of Error difference[\%] between \alname and RE-EM. Left figure: $\bbold^*\perp\bbold^*$ setting. Right Figure: $\bbold^*=\bbold^*$ setting.} \label{Figure, RE-EM}
\end{figure}

\subsection{Real Data Analysis}
Detailed information about the settings of the real data analyses is given below. 

General information:
\begin{itemize}
    \item The number of covariates that were sampled at each split in the RF implementations is $round\big(log_2(p)\big).$
    \item The number of trees that are used for the RF is not fixed. The fitting process was stooped once the overall RF error was converged in both algorithms, \alname and the standard regression tree.
\end{itemize}



Specific information for each data set analysis:
\begin{itemize}
    \item \textit{FIFA}
    \begin{itemize}[itemsep=0mm]
    \item Dependent variable: Player's market values
    \item Cluster variable: Player's club
    \item Loss function type: CV
        \item Number of covariates: $11$ 
        \item Regression tree depth: $5$
        \item Minimum number of observations in a node: $3$
        \item Number of trees for RF: $60$
        \item Comments: The covariates that are used in this analysis are: 'Age', 'Overall', 'Potential', 'Wage', 'Special', 'Preferred Foot',
       'International Reputation', 'Weak Foot', 'Skill Moves', 'Height',
       'Weight'. Other variables have many missing values or are irrelevant.
    \end{itemize}
    \item \textit{Crimes (Communities and Crime in US)}
            \begin{itemize}[itemsep=0mm]
    \item Dependent variable: Violent crimes in US communities per population size
    \item Cluster variable: State
    \item Loss function type: CV
        \item Number of covariates: $100$ (all the available covariates were used)
        \item Regression tree depth: $5$
        \item Minimum number of observations in a node: $3$
        \item Number of trees for RF: $55$
        \item Comments: - 
    \end{itemize}
    \item \textit{Korea Temperature ('bias correction of numerical prediction model temperature forecast')}
    \begin{itemize}[itemsep=0mm]
        \item Dependent variable: Daily maximum temperature at several sites in South Korea
    \item Cluster variable: Day
        \item Loss function type: Cp (since training and test set measurements are sampled from the same sites)
        \item Number of covariates: $4$
        \item Regression tree depth: $5$
        \item Minimum number of observations in a node: $3$
        \item Number of trees for RF: $70$
        \item Comments:
        \begin{itemize}
            \item  The covariates that are used in this analysis are: 'Present\_Tmin','DEM', 'Slope', 'Solar radiation'. Other variables in this data set are models's scores of the data set supplier, which are based on previous dependent variable measurements (and therefore cannot be used in LMM framework)
            \item Exponential kernel covariance function was used. Maximal temperature of different days are assumed to be uncorrelated.
            \item The original data set contains records from July and August. Due to many missing values in July along the years, only records from August are analyzed. Also, sites with missing values along the years were omitted.
        \end{itemize}  
    \end{itemize}
    \item \textit{California Housing}
        \begin{itemize}[itemsep=0mm]
    \item Dependent variable: Values of houses in California
    \item Cluster variable: Block's cluster (blocks with the same coordinate values)
        \item Loss function type: CV
        \item Number of covariates: $6$ (all the available covariates were used)
        \item Regression tree depth: $5$
        \item Minimum number of observations in a node: $3$
        \item Number of trees for RF: $50$
        \item Comments: -
    \end{itemize}
    \item \textit{Parkinson's Disease Telemonitoring}
        \begin{itemize}[itemsep=0mm]
                \item Dependent variable: Total UPDRS score, which is a score of Parkinson's Disease progression
    \item Cluster variable: Patient
        \item Loss function type: Cp (all the individuals receive approximately the same covariate values)
\item Number of covariates: $18$
        \item Regression tree depth: $5$
        \item Minimum number of observations in a node: $3$
        \item Number of trees for RF: $80$
        \item Comments: All the supplied covariates were used except the motor\_UPDRS (which its relation with the dependent variable is not fully clear to us)    \end{itemize}
    \item \textit{Wages}
        \begin{itemize}[itemsep=0mm]
       \item Dependent variable: Average hourly wages
    \item Cluster variable: Employee
    \item Loss function type: CV
        \item Number of covariates: $6$ (all the available covariates were used)
        \item Regression tree depth: $4$
        \item Minimum number of observations in a node: $3$
        \item Number of trees for RF: $80$
        \item Comments: - 
    \end{itemize}

\end{itemize}

\vskip 0.2in
\bibliography{References.bib}

\begin{thebibliography}{34}
\providecommand{\natexlab}[1]{#1}
\providecommand{\url}[1]{\texttt{#1}}
\expandafter\ifx\csname urlstyle\endcsname\relax
  \providecommand{\doi}[1]{doi: #1}\else
  \providecommand{\doi}{doi: \begingroup \urlstyle{rm}\Url}\fi

\bibitem[Akaike(1974)]{akaike1974new}
Hirotugu Akaike.
\newblock A new look at the statistical model identification.
\newblock \emph{IEEE Transactions on Automatic Control}, 19\penalty0
  (6):\penalty0 716--723, 1974.

\bibitem[Breiman(2001)]{breiman2001random}
Leo Breiman.
\newblock Random forests.
\newblock \emph{Machine learning}, 45\penalty0 (1):\penalty0 5--32, 2001.

\bibitem[Breiman et~al.(1984)Breiman, Friedman, Stone, and
  Olshen]{breiman1984classification}
Leo Breiman, Jerome Friedman, Charles~J Stone, and Richard~A Olshen.
\newblock \emph{Classification and regression trees}.
\newblock CRC press, 1984.

\bibitem[Caywood et~al.(2017)Caywood, Roberts, Colombe, Greenwald, and
  Weiland]{caywood2017gaussian}
Matthew~S Caywood, Daniel~M Roberts, Jeffrey~B Colombe, Hal~S Greenwald, and
  Monica~Z Weiland.
\newblock Gaussian process regression for predictive but interpretable machine
  learning models: An example of predicting mental workload across tasks.
\newblock \emph{Frontiers in human neuroscience}, 10:\penalty0 647, 2017.

\bibitem[Cho et~al.(2020)Cho, Yoo, Im, and Cha]{cho2020comparative}
Dongjin Cho, Cheolhee Yoo, Jungho Im, and Dong-Hyun Cha.
\newblock Comparative assessment of various machine learning-based bias
  correction methods for numerical weather prediction model forecasts of
  extreme air temperatures in urban areas.
\newblock \emph{Earth and Space Science}, 7\penalty0 (4), 2020.

\bibitem[Coull et~al.(2001)Coull, Schwartz, and Wand]{coull2001respiratory}
Brent~A Coull, Joel Schwartz, and MP~Wand.
\newblock Respiratory health and air pollution: additive mixed model analyses.
\newblock \emph{Biostatistics}, 2\penalty0 (3):\penalty0 337--349, 2001.

\bibitem[Criminisi et~al.(2011)Criminisi, Shotton, and
  Konukoglu]{criminisi2011decision}
Antonio Criminisi, Jamie Shotton, and Ender Konukoglu.
\newblock Decision forests for classification, regression, density estimation,
  manifold learning and semi-supervised learning.
\newblock \emph{Microsoft Research Cambridge, Tech. Rep. MSRTR-2011-114},
  5\penalty0 (6):\penalty0 12, 2011.

\bibitem[Fokkema et~al.(2018)Fokkema, Smits, Zeileis, Hothorn, and
  Kelderman]{fokkema2018detecting}
Marjolein Fokkema, Niels Smits, Achim Zeileis, Torsten Hothorn, and Henk
  Kelderman.
\newblock Detecting treatment-subgroup interactions in clustered data with
  generalized linear mixed-effects model trees.
\newblock \emph{Behavior research methods}, 50\penalty0 (5):\penalty0
  2016--2034, 2018.

\bibitem[Freund et~al.(1999)Freund, Schapire, and Abe]{freund1999short}
Yoav Freund, Robert Schapire, and Naoki Abe.
\newblock A short introduction to boosting.
\newblock \emph{Journal-Japanese Society For Artificial Intelligence},
  14\penalty0 (771-780):\penalty0 1612, 1999.

\bibitem[Friedman(2001)]{friedman2001greedy}
Jerome~H Friedman.
\newblock Greedy function approximation: a gradient boosting machine.
\newblock \emph{Annals of statistics}, pages 1189--1232, 2001.

\bibitem[Goovaerts(1999)]{goovaerts1999geostatistics}
Pierre Goovaerts.
\newblock Geostatistics in soil science: state-of-the-art and perspectives.
\newblock \emph{Geoderma}, 89\penalty0 (1-2):\penalty0 1--45, 1999.

\bibitem[Hajjem et~al.(2014)Hajjem, Bellavance, and Larocque]{hajjem2014mixed}
Ahlem Hajjem, Fran{\c{c}}ois Bellavance, and Denis Larocque.
\newblock Mixed-effects random forest for clustered data.
\newblock \emph{Journal of Statistical Computation and Simulation}, 84\penalty0
  (6):\penalty0 1313--1328, 2014.

\bibitem[Harville et~al.(1976)]{harville1976extension}
David Harville et~al.
\newblock Extension of the {G}auss-{M}arkov theorem to include the estimation
  of random effects.
\newblock \emph{The Annals of Statistics}, 4\penalty0 (2):\penalty0 384--395,
  1976.

\bibitem[Hastie et~al.(2009)Hastie, Tibshirani, and
  Friedman]{Hastie2009elements}
Trevor Hastie, Robert Tibshirani, and JH~Friedman.
\newblock \emph{The elements of statistical learning: data mining, inference,
  and prediction}.
\newblock New York, NY: Springer, 2009.

\bibitem[Hengl et~al.(2018)Hengl, Nussbaum, Wright, Heuvelink, and
  Gr{\"a}ler]{hengl2018random}
Tomislav Hengl, Madlene Nussbaum, Marvin~N Wright, Gerard~BM Heuvelink, and
  Benedikt Gr{\"a}ler.
\newblock Random forest as a generic framework for predictive modeling of
  spatial and spatio-temporal variables.
\newblock \emph{PeerJ}, 6, 2018.

\bibitem[Hodges and Sargent(2001)]{hodges2001counting}
James~S Hodges and Daniel~J Sargent.
\newblock Counting degrees of freedom in hierarchical and other
  richly-parameterised models.
\newblock \emph{Biometrika}, 88\penalty0 (2):\penalty0 367--379, 2001.

\bibitem[Mallows(1973)]{mallows1973some}
Colin~L Mallows.
\newblock Some comments on c p.
\newblock \emph{Technometrics}, 15\penalty0 (4):\penalty0 661--675, 1973.

\bibitem[Ngufor et~al.(2019)Ngufor, Van~Houten, Caffo, Shah, and
  McCoy]{ngufor2019mixed}
Che Ngufor, Holly Van~Houten, Brian~S Caffo, Nilay~D Shah, and Rozalina~G
  McCoy.
\newblock Mixed effect machine learning: A framework for predicting
  longitudinal change in hemoglobin a1c.
\newblock \emph{Journal of biomedical informatics}, 89:\penalty0 56--67, 2019.

\bibitem[Pace and Barry(1997)]{pace1997sparse}
R~Kelley Pace and Ronald Barry.
\newblock Sparse spatial autoregressions.
\newblock \emph{Statistics \& Probability Letters}, 33\penalty0 (3):\penalty0
  291--297, 1997.

\bibitem[Painsky and Rosset(2016)]{painsky2016cross}
Amichai Painsky and Saharon Rosset.
\newblock Cross-validated variable selection in tree-based methods improves
  predictive performance.
\newblock \emph{IEEE transactions on pattern analysis and machine
  intelligence}, 39\penalty0 (11):\penalty0 2142--2153, 2016.

\bibitem[Prokhorenkova et~al.(2017)Prokhorenkova, Gusev, Vorobev, Dorogush, and
  Gulin]{prokhorenkova2017catboost}
Liudmila Prokhorenkova, Gleb Gusev, Aleksandr Vorobev, Anna~Veronika Dorogush,
  and Andrey Gulin.
\newblock Catboost: unbiased boosting with categorical features.
\newblock \emph{arXiv preprint arXiv:1706.09516}, 2017.

\bibitem[Rabinowicz and Rosset(2020)]{rabinowicz2020cross}
Assaf Rabinowicz and Saharon Rosset.
\newblock Cross-validation for correlated data.
\newblock \emph{Journal of the American Statistical Association}, pages 1--14,
  2020.

\bibitem[Rasmussen(2003)]{rasmussen2003gaussian}
Carl~Edward Rasmussen.
\newblock Gaussian processes in machine learning.
\newblock In \emph{Summer school on machine learning}, pages 63--71. Springer,
  2003.

\bibitem[Redmond and Baveja(2002)]{redmond2002data}
Michael Redmond and Alok Baveja.
\newblock A data-driven software tool for enabling cooperative information
  sharing among police departments.
\newblock \emph{European Journal of Operational Research}, 141\penalty0
  (3):\penalty0 660--678, 2002.

\bibitem[Sela and Simonoff(2012)]{sela2012re}
Rebecca~J Sela and Jeffrey~S Simonoff.
\newblock Re-em trees: a data mining approach for longitudinal and clustered
  data.
\newblock \emph{Machine learning}, 86\penalty0 (2):\penalty0 169--207, 2012.

\bibitem[Singer et~al.(2003)Singer, Willett, Willett,
  et~al.]{singer2003applied}
Judith~D Singer, John~B Willett, John~B Willett, et~al.
\newblock \emph{Applied longitudinal data analysis: Modeling change and event
  occurrence}.
\newblock Oxford university press, 2003.

\bibitem[Speiser et~al.(2019)Speiser, Wolf, Chung, Karvellas, Koch, and
  Durkalski]{speiser2019bimm}
Jaime~Lynn Speiser, Bethany~J Wolf, Dongjun Chung, Constantine~J Karvellas,
  David~G Koch, and Valerie~L Durkalski.
\newblock Bimm forest: A random forest method for modeling clustered and
  longitudinal binary outcomes.
\newblock \emph{Chemometrics and Intelligent Laboratory Systems}, 185:\penalty0
  122--134, 2019.

\bibitem[Stephan et~al.(2015)Stephan, Stegle, and Beyer]{stephan2015random}
Johannes Stephan, Oliver Stegle, and Andreas Beyer.
\newblock A random forest approach to capture genetic effects in the presence
  of population structure.
\newblock \emph{Nature communications}, 6:\penalty0 7432, 2015.

\bibitem[Stone(1974)]{stone1974cross}
Mervyn Stone.
\newblock Cross-validatory choice and assessment of statistical predictions.
\newblock \emph{Journal of the Royal Statistical Society: Series B
  (Methodological)}, 36\penalty0 (2):\penalty0 111--133, 1974.

\bibitem[Tsanas et~al.(2009)Tsanas, Little, McSharry, and
  Ramig]{tsanas2009accurate}
Athanasios Tsanas, Max Little, Patrick McSharry, and Lorraine Ramig.
\newblock Accurate telemonitoring of parkinson’s disease progression by
  non-invasive speech tests.
\newblock \emph{Nature Precedings}, pages 1--1, 2009.

\bibitem[Vaida and Blanchard(2005)]{vaida2005conditional}
Florin Vaida and Suzette Blanchard.
\newblock Conditional akaike information for mixed-effects models.
\newblock \emph{Biometrika}, 92\penalty0 (2):\penalty0 351--370, 2005.

\bibitem[Verbeke(1997)]{verbeke1997linear}
Geert Verbeke.
\newblock Linear mixed models for longitudinal data.
\newblock In \emph{Linear mixed models in practice}, pages 63--153. Springer,
  1997.

\bibitem[Westveld et~al.(2011)Westveld, Hoff, et~al.]{westveld2011mixed}
Anton~H Westveld, Peter~D Hoff, et~al.
\newblock A mixed effects model for longitudinal relational and network data,
  with applications to international trade and conflict.
\newblock \emph{The Annals of Applied Statistics}, 5\penalty0 (2A):\penalty0
  843--872, 2011.

\bibitem[Wolfinger and O'connell(1993)]{wolfinger1993generalized}
Russ Wolfinger and Michael O'connell.
\newblock Generalized linear mixed models a pseudo-likelihood approach.
\newblock \emph{Journal of statistical Computation and Simulation}, 48\penalty0
  (3-4):\penalty0 233--243, 1993.

\end{thebibliography}

\end{document}